\documentclass[screen,sigconf]{acmart}
\usepackage{amsmath}
\usepackage{multirow}
\usepackage{caption}
\usepackage{subcaption}
\usepackage{atbegshi,picture}
\usepackage{array}
\usepackage{booktabs}
\usepackage{longtable}
\usepackage{tabularx}

\definecolor{DarkBlue}{HTML}{00008B}
\definecolor{green}{rgb}{0.0, 0.65, 0.31}
\definecolor{purple}{HTML}{ae017e}
\definecolor{mediumblue}{rgb}{0.0, 0.0, 0.8}
\definecolor{orange}{HTML}{D55E00}     
\definecolor{deeppink}{HTML}{D33682}   
\definecolor{olive}{HTML}{808000}

\newcommand{\edit}[1]{{\textcolor{black}{#1}}}

\newif{\ifhidecomments}
  \hidecommentsfalse 
\ifhidecomments
    \newcommand{\chelsea}[1]{}
    \newcommand{\hong}[1]{}
    \newcommand{\jini}[1]{}
    \newcommand{\jodi}[1]{}
    \newcommand{\alicia}[1]{}
    \newcommand{\avanita}[1]{}

\else
    \newcommand{\chelsea}[1]{\textbf{\small\sffamily{\textcolor{purple}{[#1 -- Chelsea]}}}}
    \newcommand{\hong}[1]{\textbf{\small\sffamily{\textcolor{orange}{[#1 -- Hong]}}}}
    \newcommand{\jini}[1]{\textbf{\small\sffamily{\textcolor{DarkBlue}{[#1 -- Jini]}}}}
    \newcommand{\jodi}[1]{\textbf{\small\sffamily{\textcolor{red}{[#1 -- Jodi]}}}}
    \newcommand{\alicia}[1]{\textbf{\small\sffamily{\textcolor{green}{[#1 -- Alicia]}}}}
    \newcommand{\avanita}[1]{\textbf{\small\sffamily{\textcolor{olive}{[#1 -- Avanita]}}}}
  \fi
\AtBeginDocument{%
  }

\copyrightyear{2026}
\acmYear{2026}
\setcopyright{cc}
\setcctype{by}
\acmConference[CHI '26]{Proceedings of the 2026 CHI Conference on Human Factors in Computing Systems}{April 13--17, 2026}{Barcelona, Spain}
\acmBooktitle{Proceedings of the 2026 CHI Conference on Human Factors in Computing Systems (CHI '26), April 13--17, 2026, Barcelona, Spain}
\acmPrice{}
\acmDOI{10.1145/3772318.3790936}
\acmISBN{979-8-4007-2278-3/2026/04}

\begin{document}

\title{Situated, Dynamic, and Subjective: Envisioning the Design of Theory-of-Mind-Enabled Everyday AI with Industry Practitioners}
\renewcommand{\shorttitle}{Envisioning the Design of Theory-of-Mind-Enabled Everyday AI with Industry Practitioners}

\author{Qiaosi Wang}
\orcid{0000-0002-5296-5440}
\affiliation{%
  \department{Human-Computer Interaction Institute}
  \institution{Carnegie Mellon University}
  \city{Pittsburgh}
  \state{PA}
  \country{USA}
}
\email{qiaosiw@andrew.cmu.edu}

\author{Jini Kim}
\orcid{0000-0001-8199-3240}
\affiliation{%
  \department{Human-Computer Interaction Institute}
  \institution{Carnegie Mellon University}
  \city{Pittsburgh}
  \state{PA}
  \country{USA}
}
\email{jinik@andrew.cmu.edu}

\author{Avanita Sharma}
\orcid{0009-0000-1216-0931}
\affiliation{%
  \department{School of Design}
  \institution{Carnegie Mellon University}
  \city{Pittsburgh}
  \state{PA}
  \country{USA}
}
\email{avanitas@andrew.cmu.edu}

\author{Alicia (Hyun Jin) Lee}
\orcid{0009-0006-2045-2479}
\affiliation{%
  \department{Human-Computer Interaction Institute}
  \institution{Carnegie Mellon University}
  \city{Pittsburgh}
  \state{PA}
  \country{USA}
}
\email{hlee3@andrew.cmu.edu}

\author{Jodi Forlizzi}
\orcid{0000-0002-7161-075X}
\affiliation{%
  \department{Human-Computer Interaction Institute}
  \institution{Carnegie Mellon University}
  \city{Pittsburgh}
  \state{PA}
  \country{USA}
}
\email{forlizzi@cs.cmu.edu}

\author{Hong Shen}
\orcid{0000-0002-5364-3718}
\affiliation{%
  \department{Human-Computer Interaction Institute}
  \institution{Carnegie Mellon University}
  \city{Pittsburgh}
  \state{PA}
  \country{USA}
}
\email{hongs@cs.cmu.edu}

\renewcommand{\shortauthors}{Qiaosi Wang, et al.}

\begin{abstract}
 Theory of Mind (ToM)—the ability to infer what others are thinking (e.g., intentions) from observable cues—is traditionally considered fundamental to human social interactions. This has sparked growing efforts in building and benchmarking AI’s ToM capability, yet little is known about how such capability could translate into the design and experience of everyday user-facing AI products and services. We conducted 13 co-design sessions with 26 U.S.-based AI practitioners to envision, reflect, and distill design recommendations for ToM-enabled everyday AI products and services that are both future-looking and grounded in the realities of AI design and development practices. Analysis revealed three interrelated design recommendations: ToM-enabled AI should 1) be situated in the social context that shape users' mental states, 2) be responsive to the dynamic nature of mental states, and 3) be attuned to subjective individual differences. We surface design tensions within each recommendation that reveal a broader gap between practitioners' envisioned futures of ToM-enabled AI and the realities of current AI design and development practices. These findings point toward the need to move beyond static, inference-driven approach to ToM and toward designing ToM as a pervasive capability that supports continuous human-AI interaction loops.
 
 % We discuss design implications and directions towards responsibly integrating ToM into everyday AI products and services, necessitating new approaches that center the inherent complexity of human mental states.   
\end{abstract}


% \keywords{Theory of Mind, Social Cognition Theories, Design, Socially Intelligent AI, Everyday AI Products and Services, Industry AI Practices, Storyboarding, Co-Design Workshop, Interview}


\maketitle
\section{Introduction}
From early HCI pioneers~\cite{breazeal2004designing,nass1994computers} to popular media, visions of socially intelligent AI systems that blend seamlessly into everyday interactions have been around for decades. For instance, the Kismet robot~\cite{breazeal2004designing} that can detect facial expressions and vocal tones and respond with gaze shifts or affective cues; the robots in \textit{Interstellar} that can interpret human intentions, adapt to shifting goals, and even use humor to ease stressful situations. These visions share a common aspiration--- technologies that can recognize unspoken intentions, emotions, and needs in the moment. Such capabilities reflect a socio-cognitive capacity traditionally viewed as fundamental to everyday social interactions: \textit{Theory of Mind (ToM)}, the ability to infer others' transient mental states (e.g., intentions, beliefs, desires) from observable cues~\cite{Baron-cohen1999EvolutionMind,premack1978does,baron1985does}. Some scholars have argued that ToM underlies many core human social behaviors, including anticipating actions, repairing misunderstandings, and coordinating joint plans~\cite{Baron-cohen1999EvolutionMind},
% ToM allows people to anticipate others’ actions, repair misunderstandings, and coordinate joint plans~\cite{Baron-cohen1999EvolutionMind}, 
all of which require forming inferences and conjectures about what's going on in others' minds at the moment to behave accordingly and achieve optimal social interaction outcomes.

% ToM helps us navigate everyday social interactions by inferring others' thoughts and feelings, anticipating their behavior, and communicating effectively based on observable cues. For example, we sense a roommate’s annoyance from the way they slam the door, we recognize a friend's joke and sarcasm from their playful tones and exaggerated phrasing. Many human social behaviors are enabled by ToM, such as persuasion, teaching, repairing communication breakdowns, and building shared plans and goals~\cite{Baron-cohen1999EvolutionMind}. All of these require us to make conjectures about what's going on in others' minds (e.g., their intentions, knowledge, preference, motivations) at the moment to behave accordingly and achieve optimal social interaction outcomes. 

ToM has become a growing area of interest in AI, where researchers are actively building and evaluating AI's ToM capability to improve its social adeptness. Researchers have built AI's ToM-like capability to infer human collaborator's knowledge of the environment~\cite{matarese2022perception,lo2020planning}, perception of risks~\cite{kwon2020humans}, real-time attention~\cite{huang2010joint}, or even human understanding of the AI~\cite{akula2022cx,fukuchi2022conveying} through a variety of cognitive architectures and machine learning techniques~\cite{mao2024review}. However, ToM’s largely disembodied, inference-driven framing of social cognition has faced longstanding critiques across disciplines~\cite{gallagher2004understanding,de2007participatory}. These critiques have intensified with the recent appropriations of ToM tasks designed for human children as a social intelligence benchmark for Large Language Models (LLMs)--- some AI researchers have made bold claims about LLMs ``spontaneously'' possessing human-level ToM capability~\cite{bubeck2023sparks,kosinski2023theory}, which have sparked lively debates and growing critiques \cite{ullman2023large,shapira2023clever}.
% More recently, ToM has been applied as a benchmark of Large Language Models (LLMs)' level of social intelligence, with some researchers making bold claims that human-level ToM capability has ''spontaneously emerged'' in LLMs~\cite{bubeck2023sparks,kosinski2023theory}\cite[cf.][]{ullman2023large,shapira2023clever}. With rapid advancements in AI techniques, we are closer than ever to seeing ToM-enabled AI entering everyday products and services to bring the decade-long vision into reality. 
Amid such speculative horizon around ToM development and evaluation in AI, it remains unclear how ToM-like capability in AI should be manifested and designed in everyday user-facing contexts. 

ToM-enabled AI departs from conventional AI systems that rely on explicit and static user profiles by emphasizing AI's ability to recognize and respond to humans' transient, implicit mental states. While personalization and recommendation algorithms typically predict long-term preferences from past purchases, browsing histories, or demographics~\cite{rafieian2023ai,cunha2018metalearning}, ToM-enabled AI must infer hidden, moment-to-moment states such as frustration with the AI, hesitation about a purchase, or intentions to disengage from a task. Designing for such capability presents a vast design space since human mental states are infinite, fleeting, and varied, making it infeasible to build and test all possible instantiations systematically. Design research is particularly effective in this context, as it allows us to explore and critically assess technologies before they exist, to probe potential speculative futures before social norms have stabilized around their use, and to surface considerations that engineering- or data-driven approaches may overlook. 

% This fundamental difference underscores the need to examine ToM-enabled AI as a distinct area of inquiry for AI products and services design, especially when they are likely to reshape the future of everyday human-AI interactions. 

In this paper, we envision the design and experience of futuristic ToM-enabled AI products and services in everyday contexts. We conducted 13 co-design sessions with 26 industry practitioners who have worked on user-facing AI applications, products, and services to distill design recommendations that are both future-looking and grounded in the realities of AI design and development practices. Each co-design session engaged two practitioners (one engineering-oriented, and the other design-oriented) to simulate the cross-functional industry environment. The sessions centered on learning, designing, and reflecting on ToM-enabled AI products and services across six everyday human-AI interaction scenarios. Specifically, our study examined these two research questions:
% to surface design recommendations for future ToM-enabled everyday AI products and services:

\begin{table}[h]
% \sffamily
\centering
% \footnotesize
% \vspace{-0.5em}
\begin{tabular}{@{}l@{}p{0.9\columnwidth}@{}}
    \textbf{RQ 1: } & How do AI practitioners envision and design everyday ToM-enabled AI products and services?\\
    \textbf{RQ 2: } & What are some potential challenges and opportunities in designing and developing ToM-enabled AI products and services in real-world user-facing settings? \newline
\end{tabular}
% \vspace{-0.5em}  
\end{table}

We analyzed the design artifacts through affinity diagramming and the session transcripts through reflexive thematic analysis. Our analysis revealed three interrelated design recommendations for future ToM-enabled AI products and services: ToM-enabled AI should (1) be situated in the social context that shapes people's mental states, (2) be responsive to the dynamic and moment-to-moment nature of mental states, and (3) be attuned to subjective individual differences. These recommendations captured AI practitioners' speculative visions of ToM-enabled AI, as well as critical reflections on the constraints of current AI design and development practices to realize such visions. Together, they indicate that modular, inference-driven approaches of ToM are unlikely to support the situated, dynamic, and subjective demands of everyday human-AI interaction. Building on these insights, we outline a design direction that treats ToM as a pervasive capability embedded within intended or existing AI functionalities, enabling continuous human-AI interaction loops that translate ToM-enabled AI from speculative visions to everyday products.
\section{Related Work}

\subsection{Theory of Mind, Social Cognition, and Human-AI Interaction}

Theory of Mind (ToM) was first introduced by~\citeauthor{premack1978does} [1978] as the ability to attribute mental states to oneself and others to predict behavior. Since then, ToM has been widely studied across psychology, philosophy, neuroscience, and cognitive science, and is often treated as a foundational socio-cognitive capacity that underpins children’s social development~\cite{gopnik1992child,baron1985does}, differences in social cue interpretation among autistic individuals~\cite{milton2012ontological,wellman2018theory,rakoczy2022foundations,baron2000theory}, and everyday social behaviors such as intentional communication, teaching, persuasion, and communication repair~\cite{Baron-cohen1999EvolutionMind}. Yet the definition, mechanisms, and role of ToM in social cognition remain contested. Although ToM is often framed as a cognitive process of inferring beliefs and intentions, recent work highlights its affective dimension and overlap with mechanisms of empathizing~\cite{cerniglia2019intersections,shamay2007dissociable,fu2023systematic}, both of which build on affect recognition~\cite{mitchell2015overlapping,watrin2023affect}. ToM has been commonly interpreted from the theory-theory~\cite{gopnik1992child,davies1995folk,gallagher2001practice} and simulation-theory~\cite{davies1995folk,gallese1998mirror,gallagher2001practice} perspectives that conceptualize social understanding as a cognitive, inference-based process--- either through forming folk theories of others' mental states (theory-theory) or by imaginatively simulating their perspectives (simulation-theory). These approaches have faced growing criticism for relying on a disembodied, third-person view of social cognition~\cite{plastow2012theory,gipps2004autism,gallagher2008understanding}. In contrast, interactive, embodied, and enactive approaches to social cognition argue that mental state representations are often unnecessary in everyday encounters, proposing instead that social understanding arises through direct perception, sensorimotor engagement, and participatory sense-making between individuals~\cite{gallagher2015social,de2007participatory}.

In pursuit of replicating human-level social cognition in AI systems, AI researchers have carried out these theories (and debates) into the development and evaluation of AI's social intelligence, with particular interest on AI ToM. Following largely computational and modular interpretations of ToM, prior work has developed AI's ToM cognitive architectures that model human knowledge and belief states of tasks~\cite{kwon2020humans} and infer how humans interpret AI's behaviors~\cite{fukuchi2022conveying,Devin2016AnExecution, Hiatt2011AccommodatingMind,Pynadath2005PsychSim:Agents,Harbers2009ModelingMind}.
% \edit{Building on these foundations, AI researchers have carried out these knowledge (and debates) around human social cognition to build AI's social intelligence, with particular interest and emphasis on equipping AI with ToM through computational approaches.} Researchers have built AI's ToM cognitive architecture to generate transparent and human-interpretable AI behaviors by modeling possible human interpretations of AI's motion trajectories~\cite{fukuchi2022conveying,Devin2016AnExecution, Hiatt2011AccommodatingMind,Pynadath2005PsychSim:Agents,Harbers2009ModelingMind}, or modeling human knowledge states for the AI to account for irrational human behaviors~\cite{kwon2020humans}. 
These ToM-like capabilities have been used in human-AI interactions to maximize a human collaborator's knowledge of the environment~\cite{matarese2022perception}, provide timely informational advice~\cite{lo2020planning}, improve human-AI communication outcome, engagement, and user perceptions of an AI system~\cite{buschmeier2018communicative,wagner2013developing,huang2010joint} through a variety of deep learning, reinforcement learning, and Bayesian-based modeling approaches~\cite{mao2024review}. With the rise of LLMs, researchers have also appropriated human ToM assessments (e.g., false-belief tasks, faux pas tests) to benchmark social intelligence in AI~\cite{bubeck2023sparks,kosinski2023theory}. Corresponding to the critiques and debates on ToM in social cognition literature, these ToM benchmarks have faced growing concerns and pushbacks for relying on static, synthetic, and third-person scenarios that test illusory ToM reasoning rather than capturing robust social reasoning skills~\cite{ma2023towards,ullman2023large,shapira2023clever,wang2025rethinking,hu2025re,ying2025benchmarking}. Reflecting broader shifts in social cognition, HCI scholars are increasingly advocating for interactive, embodied, and situated approaches to designing human-AI social encounters~\cite{kahl2023intertwining,deshpande2024embracing}, including work on Mutual Theory of Mind that moves beyond one-sided inference toward reciprocal, context-sensitive understanding~\cite{wang2021towards,wang2022mutual}.

Despite these advances, most AI ToM work remains centered on computational modeling and benchmark performance, offering limited insight into how ToM-like capability should be conceptualized or designed in everyday AI systems. We investigate initial design recommendations for everyday ToM-enabled AI through co-design with industry AI practitioners.

% As AI advancements are pushing AI systems to be more context- and socially-aware, ToM-like capabilities will increasingly underpin user-facing products and services. Yet it remains unclear how such capabilities should be designed when situated in everyday life. Our work investigates the initial set of design recommendations that could inform future design guidelines on everyday ToM-enabled AI products and services through co-designing with industry AI practitioners. 

% Most recently, ToM has also gained attention from the HCI community as a theoretical framework to promote human-AI mutual understanding through communication cues embedded in conversation utterances~\cite{wang2022mutual,wang2024theory,wang2021towards}.
% As AI advancements are pushing AI systems to be more context- and socially-aware, ToM-like capabilities will increasingly underpin user-facing products and services. Yet it remains unclear how such capabilities should be designed when situated in everyday life. 

\subsection{Socially Intelligent AI in Everyday Contexts}
Prior work has examined the design of socially intelligent AI systems in everyday contexts, ranging from speculative systems to real-world deployment and evaluation. These AI's social intelligence are typically enabled through personalization~\cite{reig2021social,lee2012personalization,volkel2021eliciting} or contextual awareness of the physical and social environment around the users~\cite{jaber2024cooking,lim2024exploring,yun2025if}. Personalized agents have been designed to recognize users based on prior personal histories~\cite{reig2021social,lee2012personalization}, while context-aware systems use IoT sensors or continuous monitoring devices (e.g., camera, microphone) to intervene at task breakpoints or activity transitions~\cite{oh2024better,lim2024exploring}

% For example,~\citeauthor{reig2021social} [2021] envisioned personalized robots that can recognize and greet the customer based on facial recognition and their previous visits to the auto repair shop;~\citeauthor{lee2012personalization} [2012] deployed a personalized snack robot that can make a snack delivery based on participants' service usage and prior interactions with the robot. Other work has examined AI's contextual awareness of a user's physical environment through just-in-time responses~\cite{lim2024exploring,oh2024better}. They leveraged IoT sensors to identify users' task breakpoints and activity transitions~\cite{lim2024exploring,oh2024better} or continuous monitoring devices such as cameras and microphones~\cite{luria2020social,yun2025if,jaber2024cooking} to monitor the environment. 

While these studies reported positive evaluations and envisioned futures, they also raised recurring concerns about privacy and psychological discomfort, especially in private settings. Personalized robots that act based only on user's usage and interaction with the robot in workplaces often did not trigger privacy concerns~\cite{lee2012personalization}, but some users found socially intelligent service robots ``creepy'' and were reluctant to trade personal data for friendliness or customization~\cite{reig2021social}. In smart home contexts, systems relying on cameras, microphones, or other monitoring devices raised even stronger concerns~\cite{luria2020social,lim2024exploring,chalhoub2021did}. Participants expressed discomfort even when reassured that camera sensors only detected presence without storing images~\cite{lim2024exploring}, and voiced strong negative reactions to devices that ``watch, listen, and record'' continuously, preferring instead that AI draw on existing digital traces such as emails, texts, online behavior, or even their medical records~\cite{luria2020social}.

% ~\citeauthor{lim2024exploring} [2024] pointed out that participants were still concerned about privacy risks, even though they clarified that the camera sensors only store real-time human presence without saving images. Similarly,~\citeauthor{luria2020social} [2020] also described participants' strong negative responses to continuous monitoring devices that ``watch, listen, and record'' continuously, and instead prefer the AI to leverage digital information from other sources such as their emails, texts, online behavior, or even their medical records. 

There have been growing calls for socially intelligent AI to move beyond rigid routines and rules~\cite{davidoff2006principles} to adapt to users' changing needs and social environment, which in turn shape preferences for AI behavior. For example,~\citeauthor{chang2025unremarkable} [2025] highlighted the need for AI in elderly care to evolve over time, transitioning from a tool, to a coach, to an advocate as users' cognitive capabilities decline. Other studies similarly showed that preferences for AI responses and proactiveness depend heavily on social contexts~\cite{luria2020robotic,reig2020not,luria2020social}. Users expected agents to navigate social dynamics by recognizing who was present and decide whether to engage or remain inactive~\cite{luria2020robotic}. While AI agents intervening during task transition points are considered desirable, proactive suggestions offered during ongoing tasks were sometimes experienced as especially helpful~\cite{oh2024better}.

% ~\citeauthor{luria2020robotic} [2020] described user's preferences for the AI agent to navigate social dynamics by knowing all the actors present in the situation, and also determine whether to engage or go to sleep based on the social situation.~\citeauthor{oh2024better} [2024] also pointed out that while AI agents intervening during task transitions is an established ``rule'' for desired proactive behaviors, they also found that the agent's proactive suggestions while users are engaging in specific tasks could be considered immensely helpful. 

These studies have offered valuable design insights based on user preferences and concerns~\cite{yao2023reviewing}, but they also revealed the limits of current socially intelligent AI that often rely on explicit and rigid personal and contextual information, highlighting the need for more socially sophisticated AI in everyday contexts. To move beyond these limits, our work examined the future design of ToM-enabled AI products and services with AI practitioners to provide forward-looking design recommendations that are also grounded in the realities of AI design and development practices.

\subsection{Design Innovation in AI Products and Services}
Prior work has examined the industry design innovation process, practices and associated challenges with industry AI practitioners (e.g., UX practitioners, data scientists, ML engineers). There are two major AI innovation processes: technology-centered innovation and user-centered design innovation~\cite{yang2020re,yildirim2024discovering}. Technology-centered innovation process begins with technology capabilities (e.g., AI models) or data availabilities~\cite{yang2018investigating} to continually develop and evaluate a minimal viable product (MVP)~\cite{yang2020re,yildirim2024discovering}, whereas user-centered design innovation starts from identifying user needs, preferences, and behaviors to iteratively refine and come up with usable and valuable products to their target user group~\cite{yildirim2024discovering}. Each of these processes presents their own benefits and drawbacks~\cite{yang2019sketching,yildirim2022experienced}: while technology-centered innovation can ensure product feasibility, it can sometimes constrain the design space that designers find challenging to come up with creative ideas tailored to user needs and preferences~\cite{yang2019sketching}; user-centered design innovation can maximize product usability, yet without technical constraints, designers can sometimes come up with ideas that ``cannot be built'' or focusing on problems that do not need AI~\cite{yildirim2024sketching}. 

As a result, design innovation in AI products and services remained challenging for industry teams due to a variety of factors present across both innovation paths.~\citeauthor{yang2020re} [2020] synthesized two sources of design challenges that are distinctive to AI: (1) uncertainty around AI's capabilities~\cite{dove2017ux,liao2023designerly,yang2018mapping} and (2) AI's output complexity from simple to adaptive complex~\cite{dove2017ux,subramonyam2025prototyping}. These challenges have surfaced limitations on conventional HCI design approaches: manual sketching, prototyping, or even Wizard-of-Oz methods fall short in capturing AI's infinite possible outputs when systems are adaptively complex~\cite{yang2020re,yang2019sketching}.~\citeauthor{yildirim2023creating} [2023] have proposed design resources to help designers familiarize with AI capabilities and examples prior to ideation of situations where these capabilities could be helpful. This approach blended the strengths of user-centered design and technology-centered design that resulted in a broader problem-opportunity space~\cite{yildirim2023creating}, adding to the need of moving beyond user-centered design~\cite{yildirim2023creating,forlizzi2018moving}. In addition to innovation challenges, other work has surfaced additional business and organizational pressure on proioritizing market speed and profit~\cite{wang2023designing,yildirim2022experienced}, which could lead to AI product teams skipping prototype evaluation and focus on ideal user journeys that overlook potential AI failures and responsible AI issues~\cite{moore2023failurenotes}. 

Prior work has established a foundation for understanding AI design practices and challenges, while highlighting the need for new processes to envision future systems. We build on this by engaging AI practitioners in co-design to identify design recommendations for ToM-enabled products and services and reflect on the challenges of bringing them into everyday user-facing contexts.

\section{Method}
The goal of this study is to envision how ToM-enabled AI might function in everyday products and services, and to surface the potential challenges and design opportunities involved through AI practitioners' perspectives informed by their current AI design and development practices in industry. To do this, we conducted 13 two-hour virtual co-design sessions with industry AI practitioners (n=26) to learn about, design, and reflect on everyday AI products and services that could recognize and respond to users' mental states. Given that ToM’s definition and mechanisms remain contested, we adopted a traditional definition of ToM as computational, inference-based social reasoning~\cite{gopnik1992child,baron1985does,Baron-cohen1999EvolutionMind} throughout our study. This definition, which also underlies much of current AI ToM research and development, served as a starting point for exploring how ToM might be designed in everyday AI systems. Each co-design session was conducted with two industry practitioners, one with an engineering-focused background and the other with a design-focused background. This pairing was to mimic real-world cross-functional AI design and development processes, where both engineering and design perspectives are critical in envisioning user-facing AI products and services.

There were three parts in each co-design session: (1)~\textbf{Learning about ToM}, where practitioners learned about the traditional inference-based definition and applications of ToM in human-human and human-AI interactions through interactive examples and brief discussions, (2)~\textbf{Designing ToM-enabled AI features}, where practitioners engaged in three design activities to collaborate on the design of ToM-enabled AI features that can better align AI behaviors with the user's mental states in one out of the six human-AI social misalignment scenarios prepared by the research team (described in details in section~\ref{scenario_generation}), (3)~\textbf{Reflecting on ToM-enabled AI design}, where practitioners engaged in semi-structured focus group interviews to reflect on their co-design experience and product of ToM-enabled AI in the broader context of current AI design and development practices in industry. We asked participants to envision ToM-enabled AI features speculatively during the design process, while grounding their reflections in the realities of industry AI practices during the interview. This study was approved by the IRB (Institutional Review Board) at researchers' institution.

\subsection{Study Preparation: Human-AI Social Misalignment Scenario Generation} \label{scenario_generation}
% \edit{Taking the classic inference-based view of ToM to social cognition~\cite{Baron-cohen1999EvolutionMind,baron1985does}, we envision ToM-enabled AI to potentially perform a wide range of social functionalities and features in everyday AI products and services by recognizing and responding to people's mental states. Therefore,} we iteratively prepared a series of scenarios where \edit{AI systems behave in a socially misaligned way that fails to recognize or respond to people's transient mental states during interactions.} These scenarios were used as probes to prompt practitioners to consider concrete ToM-enabled AI features that could prevent or address such \edit{social misalignments with people's mental states}. 
% \edit{Taking a classic inference-based view of ToM to social cognition~\cite{Baron-cohen1999EvolutionMind,baron1985does}, we framed ToM-enabled AI as systems capable of recognizing and responding to people’s mental states in everyday interactions. 
Based on the traditional inference-based ToM framing~\cite{baron1985does,gopnik1992child}, we framed ToM-enabled AI as systems capable of recognizing and responding to people’s mental states in everyday interactions. We iteratively developed scenarios depicting AI systems that fail to do so, behaving in socially misaligned ways that overlook or misread users’ transient mental states. These scenarios served as probes to prompt practitioners to envision concrete ToM-enabled features that could prevent or address such social misalignments with people's mental states.
To ensure that these scenarios are familiar to AI practitioners working across various user-facing AI products and services, and are representative of common everyday AI usage, we first reviewed relevant research reports from accredited and accessible sources~\cite{pew2023_ai_awareness,aiprm_ai_statistics,pega_ai_survey,gallup2025_ai_everyday_products}, most of which focused on U.S.-based consumers' everyday AI usage. Through these reports, we identified seven common everyday AI products and services (smart home devices, chatbots, recommendation algorithms, digital/virtual assistants, navigation, generative AI applications, autonomous driving cars) and eight everyday AI usage contexts (transportation and shopping, managing finances, meal preparation, house cleaning and home maintenance, managing communication with others, managing health and wellbeing, plan travel itinerary, prepare for a job interview). 

\begin{figure*}
    \centering
    \includegraphics[width=\linewidth]{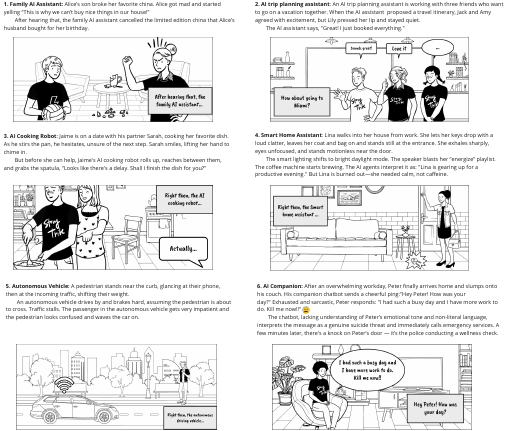}
    \caption{The six human-AI social misalignment scenarios practitioners worked on during their co-design sessions. Illustrations under each scenario description were drawn by the research team (created with StoryTribe.com) and served as the first frames of the four-frame storyboards practitioners worked on during the storyboarding activity.}
    \label{fig:hai_scenarios}
\end{figure*}

We then facilitated a one-hour brainstorming session among the research team (U.S.-based) to come up with specific human-AI social misalignment scenarios where AI did not recognize and/or respond to human's transient mental states based on the seven mental-state dimensions summarized in~\citeauthor{beaudoin2020systematic} [2022]: intentions, beliefs, understanding non-literal meanings (e.g., sarcasm), desires, emotions, knowledge, and perspectives. The research team was introduced to definitions and examples of inference-based ToM in human-human and human-AI interactions, and was encouraged to draw inspirations from personal interactions and experience with other people, everyday technologies, or envisioned AI systems. We collectively generated 127 preliminary scenarios at the end of this session. After removing similar/duplicate or incomplete scenarios, we reviewed each remaining scenario by labeling the mental-state dimensions involved (e.g., intention, knowledge) and the AI products and usage contexts described (e.g., smart home devices, meal preparation). We then evaluated the scenarios based on whether the mental states described were transient and reasonably inferrable from the scenario, and whether the scenario descriptions were clear and comprehensible. This process yielded 26 valid scenarios, from which we collectively voted to select eight scenarios that covered most of the seven ToM dimensions and represented a diverse set of AI applications and usage contexts. These scenarios were then tested through six pilot sessions and further refined and narrowed down to six scenarios based on comprehensiveness to the pilot participants. These scenarios are described below and shown in Fig.~\ref{fig:hai_scenarios}. The illustrations served as a visual cues to help practitioners better comprehend the scenarios as well as the first frames of the storyboards practitioners worked on.

\begin{enumerate}
    \item \textbf{Family AI Assistant.} Alice's son broke her favorite china. Alice got mad and started yelling ``This is why we can’t have nice things in our house!'' After hearing that, the family AI assistant canceled the limited edition china that Alice’s husband bought for her upcoming birthday.
    \item \textbf{AI Trip Planning Assistant.} An AI trip planning assistant is working with three friends who want to go on a vacation together. When the AI assistant proposes a travel itinerary, Jack and Amy agree with excitement, but Lily stays quiet. The AI assistant says, ``Great! I just booked everything.''
    \item \textbf{AI Cooking Robot.} Jaime is on a date with his partner Sarah, cooking her favorite dish. As he stirs the pan, he hesitates, unsure of the next step. Sarah smiles, lifting her hand to chime in. But before she can help, Jaime's AI cooking robot rolls up, reaches between them, and grabs the spatula, ``Looks like there's a delay. Shall I finish the dish for you?''
    \item \textbf{Smart Home Assistant.} Lina walks into her house from work. She lets her keys drop with a loud clatter, leaves her coat and bag on and stands still at the entrance. She exhales sharply, eyes unfocused, and stands motionless near the door. The smart lighting shifts to bright daylight mode. The speaker blasts her ``energize'' playlist. The coffee machine starts brewing. The AI agents interpret it as: ``Lina is gearing up for a productive evening.'' But Lina is burned out. She needed calm, not caffeine.
    \item \textbf{Autonomous Vehicle.} A pedestrian stands near the curb, glancing at their phone, then at the incoming traffic, shifting their weight. An autonomous vehicle drives by and brakes hard, assuming the pedestrian is about to cross. Traffic stalls. The passenger in the autonomous vehicle gets very impatient and the pedestrian looks confused and waves the car on.
    \item \textbf{AI Companion.} After an overwhelming workday, Peter finally arrives home and slumps onto his couch. His companion chatbot sends a cheerful ping:``Hey Peter! How was your day?'' Exhausted and sarcastic, Peter responds: ``I had such a busy day and I have more work to do. Kill me now!!'' The chatbot, lacking understanding of Peter’s emotional tone and non-literal language, interprets the message as a genuine suicide threat and immediately calls emergency services. A few minutes later, there's a knock on Peter’s door. It’s the police conducting a wellness check.
\end{enumerate}

\subsection{Participants \& Recruitment}
To identify industry practitioners who have worked on user-facing AI applications, products, and services, we distributed recruitment messages on social media (e.g., X, LinkedIn, Facebook) and personal networks, and conducted snowball recruitment through our participants. We recruited U.S.-based industry practitioners with at least six months of full-time experience designing and/or developing user-facing AI applications, products, features, or services. Participants signed up through our screening survey link included in the recruitment message. Our research team then paired up eligible participants and reached out for further scheduling.

We conducted 13 co-design sessions with 26 participants. Our participants occupied a variety of engineering and design roles, with a median of 4.5 years of experience. Half of our participants currently worked at organizations with 100,000+ corporate employees. Our participants had experience working on a wide range of AI product domains. Details of participant information can be found in table~\ref{tab:participant_info}. Each participant was compensated with a \$100 Amazon e-gift card upon successful completion of the study. 

\setcounter{table}{0}

\begin{table*}[]
    \centering
    \begin{tabular}{clclcccl} \toprule
    % \begin{tabular}{M{3cm}|M{1.5cm}|M{1.5cm}|M{1.5cm}|M{2cm}|M{4.5cm}|}
     \textbf{Session} & \textbf{Scenario} & \textbf{ID} & \textbf{Job Role}  &  \textbf{Gender}  &   \textbf{YoE}  &   \textbf{Org Size} & {\textbf{AI Product Domain}}\\ \midrule
     \multirow{2}{1em}{\centering S1} & \multirow{2}{7em}{AI Companion} &   
                      E1    &  Applied Scientist  &    M    &    5    &  100,001+    &   GenAI assistant \\
               &  &   D1    &  Product Designer   &    M    &    5    &   10,001+    &   AI-powered ads \\ \midrule
    \multirow{2}{1em}{\centering S2} & \multirow{2}{7em}{AI Trip Planning Assistant} &   
                      E2    &   Research Scientist  &    M    &    2    &    100,001+   &   GenAI tools \\
               & &    D2    &   UX Researcher       &    W    &    2    &    10,001+    & AI chatbot in education\\ \midrule
    \multirow{2}{1em}{\centering S3} & \multirow{2}{7em}{AI Companion} &   
                      E3    &  Software Engineer  &    M    &    5    &  100,001+    &   AI for bioinformatics \\
               &  &   D3    &  AI Product Designer   &    W    &    5    &   200-300    &   GenAI LLM products \\ \midrule
    \multirow{2}{1em}{\centering S4} & \multirow{2}{7em}{Autonomous Vehicle} &   
                      E4    &  Software Engineer  &    W    &    4    &  100,001+    &   AI-powered search tools \\
               &  &   D4    &  UX Researcher   &    W    &    5    &   100,001+    &   AI-powered search tools \\ \midrule
    \multirow{2}{1em}{\centering S5} & \multirow{2}{7em}{Smart Home Assistant} &   
                      E5    &  Software Engineer  &    W    &    2    &  100,001+    &   GenAI assistant \\
               &  &   D5    &  UX Researcher   &    W    &    2    &   10,001+    &   GenAI tools \& chatbot \\ \midrule
    \multirow{2}{1em}{\centering S6} & \multirow{2}{7em}{AI Trip Planning Assistant} &   
                      E6    &  ML Engineer  &    NB    &    5    &  10,001+    &   AI for medical imaging \\
               &  &   D6    &  UX Researcher  &    W    &    3    &   100,001+    &   AI design tool \\ \midrule
    \multirow{2}{1em}{\centering S7} & \multirow{2}{7em}{AI Cooking Robot} &   
                      E7    &  Application Analyst  &    M    &    10    &  10,001+    &   AI for medical imaging \\
               &  &   D7    &  UX Designer          &    W    &    2    &   100,001+    &   GenAI tools \\ \midrule
    \multirow{2}{1em}{\centering S8} & \multirow{2}{7em}{AI Trip Planning Assistant} &   
                      E8    &  GenAI Director  &    M    &    25    &  10,001+    &   Enterprise chat tools \\
               &  &   D8    &  UX Designer     &    W    &    7    &   100,001+    &   Compliance tools \\ \midrule
    \multirow{2}{1em}{\centering S9} & \multirow{2}{7em}{Family AI Assistant} &   
                      E9    &  Data Scientist  &    M    &    4    &  100,001+    &   Supply chain forecasting \\
               &  &   D9    &  UX Researcher     &    W    &    1    &   10,001+    &   Conversational AI assistant \\ \midrule
    \multirow{2}{1em}{\centering S10} & \multirow{2}{7em}{Smart Home Assistant} &   
                      E10    &  ML Engineer  &    M    &    3    &  201-500    &   AI-powered pricing system \\
               &  &   D10    &  Design Lead     &    W    &    7    &   10,001+    &   AI lifecycle platform \\ \midrule
    \multirow{2}{1em}{\centering S11} & \multirow{2}{7em}{Autonomous Vehicle} &   
                      E11    &  Research Engineer  &    M    &    5    &  1,001-5,000    &  LLM for data classification  \\
               &  &   D11    &  UX Researcher     &    M    &    7    &   10,001+    &  AI model design \& demo  \\ \midrule
    \multirow{2}{1em}{\centering S12} & \multirow{2}{7em}{Smart Home Assistant} &   
                      E12    &  AI/ML Architect  &    M    &    10    &  100,001+    &  LLM agents with customers  \\
               &  &   D12    &  Research Scientist     &    W    &    2    &   10,001+    & Conversational AI  \\ \midrule
    \multirow{2}{1em}{\centering S13} & \multirow{2}{7em}{Family AI Assistant} &   
                      E13    &  Design Engineer  &    W    &    4    &  100,001+    &  Robotic navigation  \\
               &  &   D13    &  UX Researcher &    W    &    1    &   1,001-5,000    & GenAI tool  \\
    % \multirow{5}{4em}{\centering Session 3} & \multirow{3}{4em}{\centering T3} &   P7    &   M     &    29    &    India    &    7 \\
    %     & &   P9    &   F   &    27    &    United States    &    2 \\
    %     & &   P10    &   M   &    29    &    United States    &    4 \\  \addlinespace
    %     & \multirow{2}{4em}{\centering T4} &   P8    &   M   &    31    &    England    &    4 \\
    %     & &   P11    &   F   &    27    &    United States    &    6 \\
    \bottomrule
       
    \end{tabular}
    \caption{Study participant information. ''M'' stands for ''Man'', ''W'' stands for ''Woman'', ``NB'' stands for ``Non-Binary''; ``YoE'' stands for ``Year of Experience.'' Each session consists of one practitioner in an engineering-oriented role (e.g., data scientist, software engineer, etc.) with participant ID beginning with ``E'', and one practitioner in a design-oriented role (e.g., UX researcher, product designer, etc.) with participant ID beginning with ``D''. Participants often have multiple experiences across different AI product domain, listed are the AI products that they most recently worked on. }
    \label{tab:participant_info}
\vspace*{-5pt}
\end{table*}

\subsection{Co-Design Session Procedure}
Each co-design session began with the researcher introducing the ToM concept, study motivation, and an ice-breaker activity. The researcher then proceeded with the following three parts of the session. The slide deck that illustrated the entire session procedure is available in the supplementary material. 
% All sessions were audio and video recorded. 

\subsubsection{Learning about ToM (20 min)}
We began by presenting examples of ToM-enabled behavior in human-human social interactions in detecting sarcasm, recognizing desire, and recognizing intention. These examples spanned verbal and non-verbal behaviors and across individual and group contexts in everyday life, emphasizing the inference-based approach to social understanding that underlies human social behaviors. We then asked participants to come up with examples of ToM usage in their day-to-day lives as a way for us to check and correct participants' understanding of ToM.

Next, as a warm-up activity to the main design activity, we presented two examples of human-AI interaction in everyday life (AI shopping assistant and AI study buddy) and asked participants to come up with ideas on how equipping these AI systems with ToM could cater to user's mental states in each case. All examples and procedures were tested and refined through pilot sessions to maximize comprehensiveness. 

\subsubsection{Designing ToM-enabled AI Features (50 min)}
In the design activity, practitioners collectively chose one human-AI social misalignment scenario (as shown in Fig.~\ref{fig:hai_scenarios}) to work with for the rest of the design activity. Participants were instructed to be creative, speculative, and ignore real-world constraints to surface futuristic AI features as well as potential issues and challenges. Participants went through three design activities using a combination of the design worksheet we prepared on the \href{https://www.miro.com}{Miro} virtual white-boarding tool and the \href{https://storytribeapp.com}{StoryTribe} online storyboarding tool. 

\vspace{1mm}
\noindent \textbf{1. Brainstorming Mental States.} Participants spent five minutes discussing and brainstorming the possible human mental states that could be present in the social misalignment scenario they chose. The seven dimensions of mental states from~\citeauthor{beaudoin2020systematic}[2020] were presented as an inspiration on the worksheet, but participants were encouraged to think beyond these dimensions. For example, some mental states that participants came up with for the AI trip planning assistant scenario includes: ``Lily doesn't want to go on this trip and felt forced into it by her friends.''
% ``Lily doesn't want to use an AI travel planning assistant,'' 

\vspace{1mm}
\noindent \textbf{2. Brainstorming AI Techniques.}
Second, participants chose two to three mental states that they were interested to explore further, and spent a total of 10 minutes brainstorming all possible ways that the AI could recognize each mental state. To better facilitate discussions among the participants, the researcher wrote down the ideas participants generated on the design worksheet in these two brainstorming steps. For example, some AI techniques that participants came up with to recognize Lily's mental states in the AI trip planning assistant scenario includes ``AI can monitor Lily's facial expressions and body languages.''
% , ``AI can collect social media postings and messages with friends.''

\vspace{1mm}
\noindent \textbf{3. Storyboarding.}
Finally, participants spent 20 minutes creating a four-frame storyboard that illustrated how the ToM-enabled AI can now recognize and act on the human user's mental states to prevent social misalignments. Participants used the \href{https://storytribeapp.com}{StoryTribe} online storyboarding tool for this activity. We provided instructions for each frame to better scaffold the storyboarding process, where the first frame illustrated the original scene, the second frame described AI collecting relevant information, the second frame describing how the AI can infer the correct mental states, and the fourth frame describing AI offering solutions based on the correct mental state inferences drawn. The first frame of all scenarios were already drawn out by the research team on StoryTribe to better facilitate the virtual storyboarding process.

% The first frame of the storyboard illustrated the original scene prior to the AI's actions based on misalignment of user's mental states. The first frame of all scenarios were already drawn out by the research team on StoryTribe to better facilitate the virtual storyboarding process, providing some basic materials (e.g., characters, background scene) for participants to work with in the following frames. We intentionally did not provide the visuals for the AI in the scenario and encouraged participants to illustrate the AI in their own drawings. The second frame describes the AI collecting relevant information and participants were asked to illustrate where the relevant information is presented and how is it collected by the AI. The third frame asked the participants to illustrate how the AI can analyze and infer the correct mental states of the human user. The fourth frame asks participants to illustrate AI offering solutions based on the correct mental state inference drawn. 

\subsubsection{Reflecting on ToM-enabled AI Design (40 min)}
At the end of the session, participants engaged in a semi-structured focus group interview to reflect on their designs and real-world AI practices in industry. Participants were first asked about their likes, dislikes, opinions of the ToM-enabled AI features they created, as well as similarities and differences compared to the AI products they have worked on at their jobs. They then reflected on the concept of ToM-enabled AI products and discussed things that were inspiring or difficult from their perspectives. Finally, participants were asked to envision potential challenges in designing and developing ToM-enabled AI products in current AI design and development practices as well as opportunities and contexts that would be suitable to have ToM-enabled AI. 

\subsection{Data Analysis}
All co-design sessions were audio and video recorded, and later transcribed. The design artifacts (e.g., sticky notes and storyboards created by the participants) were also collected and consolidated for data analysis. Given the variety of data generated from our co-design sessions, we used a combination of affinity diagramming~\cite{holtzblatt1997contextual} and reflexive thematic analysis~\cite{braun2021can,braun2021one} to analyze our data. 

To infer emerging design patterns, we used affinity diagramming for artifact analysis to synthesize the types of mental states that practitioners considered, AI techniques to identify human mental states, and the types of AI solutions tailored to human mental states. Two researchers evenly divided the design artifacts organized by scenarios, independently analyzed all the artifacts and grouping them into common themes, then reviewed, discussed, and consolidated these themes across the scenarios to identify patterns. This resulted in seven broad themes of mental states and 20 subthemes that categorized these mental states on a more detailed level such as ``emotional mental states because of the situation.'' For the AI techniques, we synthesized two broad categories of what data the AI can collect and how the AI can make inferences then synthesized seven and 10 categories each. Example themes and data from our affinity diagramming can be found in Appendix~\ref{affinity_analysis}.

% From this work, we synthesized seven broad themes that largely aligned with the seven dimensions of mental states identified by~\citeauthor{beaudoin2020systematic} [2020] such as emotional states, intention, desire. We also synthesized 20 sub-themes that categorized these mental states on a more detailed level such as ``emotional mental states because of the situation'' and ``intentions about the future.'' For the AI technique sticky notes, we synthesized two broad categories of what data the AI can collect and how the AI can make inferences. Under what data the AI can collect, we synthesized seven categories that covered data from a variety of modalities (e.g., body movement, digital footprint/information); under how the AI can make inferences, we synthesized 10 categories that covered different inference techniques (e.g., AI inferring social patterns, AI interpreting relationships between people). 

To understand AI practitioners' design decisions and perspectives on user-facing ToM-enabled AI products and services, we used Reflexive Thematic Analysis (RTA)~\cite{braun2021can,braun2021one} to analyze the co-design session transcripts, including discussions throughout the design activity that were not captured by the design artifacts. RTA encourages researchers to embrace their subjectivity to actively interpret, shape, and generate themes. Five researchers were involved in the RTA process and each session transcript was reviewed and coded by two researchers. We followed the six phases of RTA outlined in~\citeauthor{braun2006using} [2006]. Throughout this process, we discussed and exchanged insights during the weekly meetings and iteratively generated, searched, reviewed, and refined themes. This process resulted in 11 themes (e.g., ``ToM-enabled AI collecting data to infer mental states'', ``technical challenges in ensuring the accuracy of ToM inferences'')
% , ``technical challenges of designing and building ToM-enabled AI'', ``benefits and opportunities of ToM-enabled Ai'') 
and a total of 65 codes (e.g., ``ToM-enabled AI can collect data through network of existing devices'', ``challenging to imagine future possibility beyond available technology and constraints''). 
% , ``ToM-enabled AI can improve UX of AI systems''). 
After further discussions and review of these themes and codes, we distilled three unique design considerations of ToM-enabled AI products and services that emerged across these themes, which we present in the Findings section. We detailed the mapping between design recommendations, themes, and representative example codes in Appendix~\ref{rta}.

\subsection{Positionality Statement}
Our U.S.-based author team consists of researchers with academic research experience on design, human-AI interaction, responsible AI, and varied experience and knowledge about industry AI product design and development practices. Three authors have research experiences on designing socially intelligent AI products and services with varied populations. Four authors have conducted research on industry responsible AI practices with industry AI practitioners. All authors are currently living in the U.S. and have received bulk of their research training and/or education in predominantly Western institutions. Our background and experiences influence our positionality and shaped the subjectivity inherent in our research questions, study design, participant recruitment, and data interpretation and analysis. 
\section{Findings}
Our co-design sessions revealed three interrelated design recommendations for future everyday ToM-enabled AI products and services: ToM-enabled AI should be situated in the social context, support dynamic mental states, and attune for individual subjectivity. These design recommendations not only captured practitioners' speculative visions of what ToM-enabled AI could become, but also their critical reflections on the constraint of current AI design and development practices.
% , and user's role in shaping the capability of this technology. 
In doing so, they point toward improved futures for ToM-enabled AI that move beyond today’s systems that largely rely on static profiles or narrow input signals rather than situated, dynamic, and subjective mental states. 

% \chelsea{Add transition here.} 
% - We surfaced three distinct design recommendations for everyday ToM-enabled AI: situated, dynamic, and subjective. 
% - These design recommendations distinguished ToM-enabled AI from the current mainstream user-facing AI applications, necessitating new design and development approaches for AI systems that can naturally adapt to the user's mental states. 

% D10 (01:33:26):
% Maybe this might be a bit of an exaggeration maybe, but I'm just speaking out here. I think at some level the three of mine capabilities should apply for almost all the AI augmented features and interfaces. Because if it's just making recommendations and just doing things that doesn't necessarily have to be an AI based task. For example, I can give you an example. For example, when you're chatting with an AI chat interface, it gives you a response. Understanding to a level about the user's state of mind could help the AI chat interface respond better as opposed to giving you a regular response that it gives to every thousands and millions of people that it gives. So at that level, I think at some level it can influence or the influence is required for it to be more of a smart AI feature.

\subsection{Designing ToM-enabled AI that is Situated in the Social Context}
Our analysis showed that practitioners envisioned ToM-enabled AI as capable of inferring and acting on people's mental states in relation to the social contexts that they are situated in, rather than treating those states in isolation. This vision emerged through our analysis of practitioners' designs and reflections throughout the sessions, which revealed their emphasis on multi-modal approaches to mental state inference, unobtrusive embedding into everyday environments and infrastructures, and alignment of AI actions with social norms and situational contexts.

\begin{figure*}
    \centering
    \includegraphics[width=0.95\linewidth]{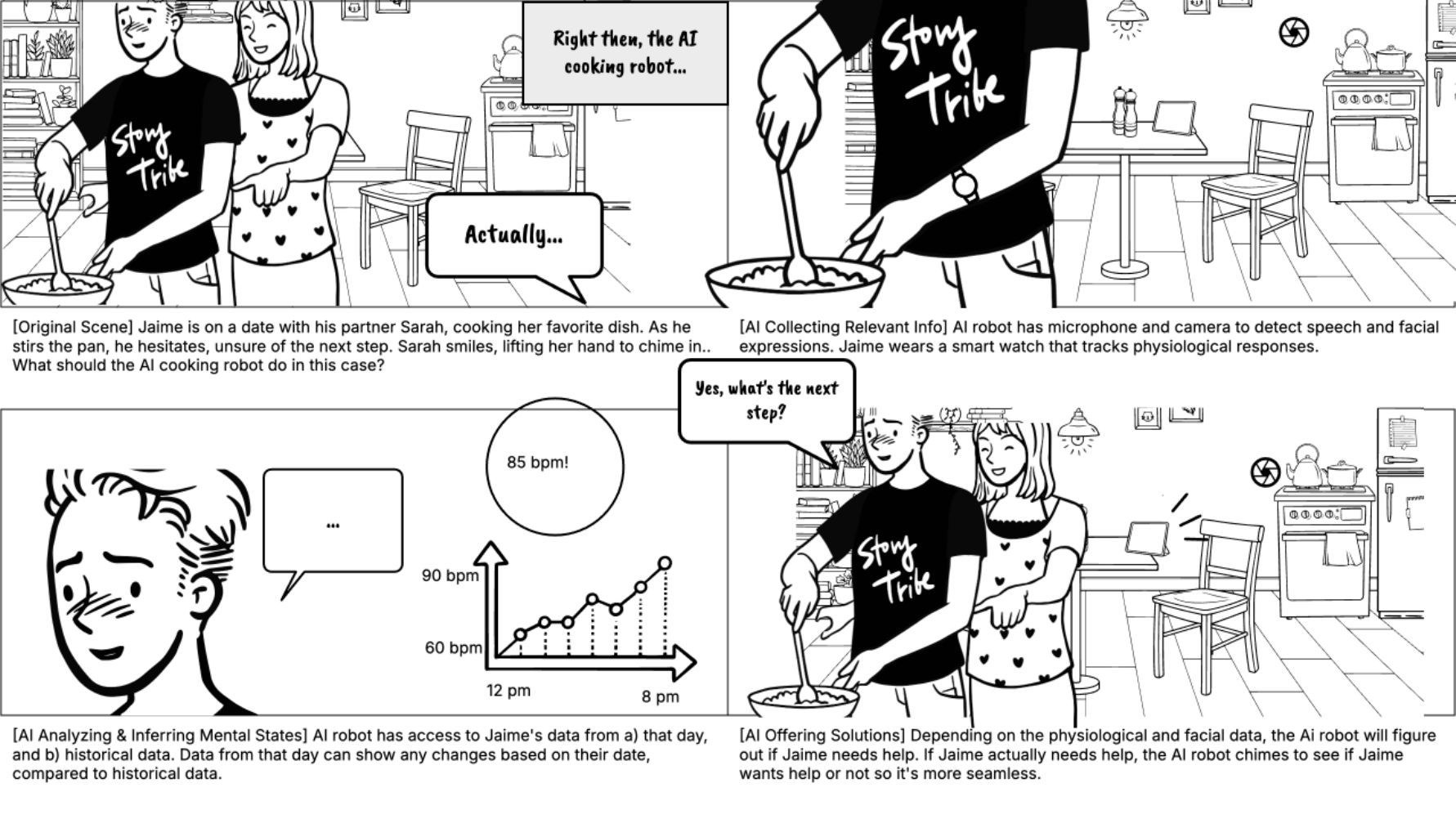}
    \caption{S7 storyboard on AI Cooking Robot. Created with StoryTribe.com.}
    \label{fig:cookingrobot_storyboard}
\end{figure*}

\subsubsection{Embracing Multi-Modal Perspectives to Infer Mental States}
Throughout the co-design sessions, we found that many practitioners embraced a multi-modal perspective in inferring human mental states by collecting data from various modalities of data sources. We summarized eight categories of data sources that practitioners came up with during the AI technique brainstorming activity across all sessions: visual cues (e.g., facial expressions), behavioral cues (e.g., body language), voice and speech cues (e.g., tone of voice), physiological and biometric information (e.g., heart rate), environmental cues (e.g., unusual noises), personal data (e.g., health records), and digital footprint (e.g., social media data). Across all the storyboards produced in the design activity, our analysis showed that practitioners typically combined two to five of the eight data sources described above to infer characters’ mental states within their scenarios. For example, S7 (AI Cooking Robot) collected the characters’ speech data, facial expressions, and physiological responses to make inferences about the character’s need for help from the AI cooking robot (as shown in Fig.~\ref{fig:cookingrobot_storyboard}). S5 (Smart Home Assistant) and S1 (AI Companion) gave their ToM-enabled AIs the ability to see the character’s home layout (e.g., messiness represented through clothes laying on the ground, uncleaned dishes laying around), behavioral cues (e.g., drinking or smoking), and facial expressions to infer the character’s desires and emotional states in the scenarios. S11 (Autonomous Vehicle) talked about accessing contextual and environmental data (e.g., nearby crosswalks, shops across from the road, GPS information) in addition to collecting pedestrian’s phone data, facial expressions, and standing postures to infer pedestrian’s intention on crossing the road.

We found that practitioners viewed this multi-modal perspective as an opportunity to capture richer signals of human mental states. As E8 reflected during the interview:
\begin{quote}
\textit{``Going through this design exercise made me realize how little we get from just words. It reminds me of when you text someone, it's very difficult sometimes to determine someone's tone. People make a lot of assumptions because we're missing a visual or even an audio. And this whole thing is making me think, there's an opportunity here. We are going to need to have more sensors to be able to grab these nonverbal types of things.''} (E8)
\end{quote}
In addition to collecting multi-modality data from the individual users, practitioners highlighted that another opportunity could be to combine multi-modality personal data with multi-modality environmental cues. D10 said,~\textit{``Combining the [character's] breathing patterns, with the location, the time could also say things. So the inference comes from multiple data points from the person combined together to predict what that person is feeling.''}

However, practitioners also pointed out that handling this amount of multi-modality data from different sources could be challenging given the current AI development paradigm and data infrastructure. E9 highlighted the significant computational power required just to collect and clean all the data required: \textit{``[The technical challenge] is not only the collection of the data, it's the standardization, the formatting of the data, typical machine learning type issues. And you're going to have a huge, huge data set with hundreds, thousands beyond that in terms of potential input for parameters and features and things. I mean, that's a lot of computational power to build a ToM-enabled AI.''} E2 echoed this point and expressed concerns about designing such architecture to also support privacy:~\textit{``I feel like there is a lot more data that are involved in this process and it's very tricky to design an architecture that supports this massive streams of data and at the same time support privacy by design.''}

% Throughout the study session, participants recognized that human mental states are often implicit, and can be predicted through various modalities of behavioral, verbal, and environmental cues. This realization has pushed participants to embrace a multi-modality perspective in inferring human mental states when designing ToM-enabled AI. For example, E8 reflected during the interview:~\textit{“Going through this design exercise made me realize how little we get from just words. It reminds me of when you text someone, it's very difficult sometimes to determine someone's tone. People make a lot of assumptions because we're missing a visual or even an audio. And this whole thing is making me think, there's an opportunity here. We are going to need to have more sensors to be able to grab these nonverbal types of things.”} In addition to multi-modality data from the individual users, practitioners also reflected on the need to collect and combine multi-modality personal data with environmental cues that the person is situated in. D10 said,~\textit{“I could also see what E10 is saying in terms of the time of the day. For example, combining it with the breathing patterns, like the location, the time and breathing patterns could also say things. So the inference comes from multiple data points from the person combined together to predict what that person is feeling.”}

\subsubsection{Designing for Situatedness Through Unremarkable Infrastructure}
Across all sessions, we found that practitioners spent great efforts in designing ToM-enabled AI to be unremarkable or even invisible to blend into the interaction environment. To our surprise, outside of the autonomous vehicle scenario, only S1 (AI Companion) designed a more overt and embodied AI form--- a physical robot that could dance and tell jokes. In the AI cooking robot scenario where the AI was clearly described as a robot in the physical form, S7 still transformed it into a small tablet that sits quietly on the desk (as shown in Fig.~\ref{fig:cookingrobot_storyboard}). In the Smart Home Assistant scenario storyboards, the AI was illustrated as a photo frame in the background with a view of the entire room (S5), or as an ambient, cylinder-shaped speaker device similar to Amazon’s Echo (S10). Practitioners who worked on scenarios that are not home-based, such as the AI Trip Planning Assistant scenario, chose to not have a physical form of the AI and represented the AI as \textit{``living on the cloud''} (S2), as a virtual agent (S8), or as a traditional desktop computer (S6). 

% However, in order to capture multi-modality data such as the character’s facial expression, S6 participants felt compelled to add certain camera equipment in their storyboard. They articulated their concerns about the unnaturalness this brought to the scene during storyboarding:
% \begin{quote}
% E6: \textit{I don't know [about this camera tripod illustration]. Let's just try that.} \\
% D6: \textit{Yeah, it's just [to illustrate that] there's something watching them and it happens that they're all in the screen. But that's a limitation of this is they have to know how to position their bodies so that the AI can help them the best. And so it's like one of them can't be facing away with a side face. They have to all be looking at it, which is also a little bit unnatural when three friends are talking.}
% \end{quote}

\begin{figure*}
    \centering
    \includegraphics[width=0.95\linewidth]{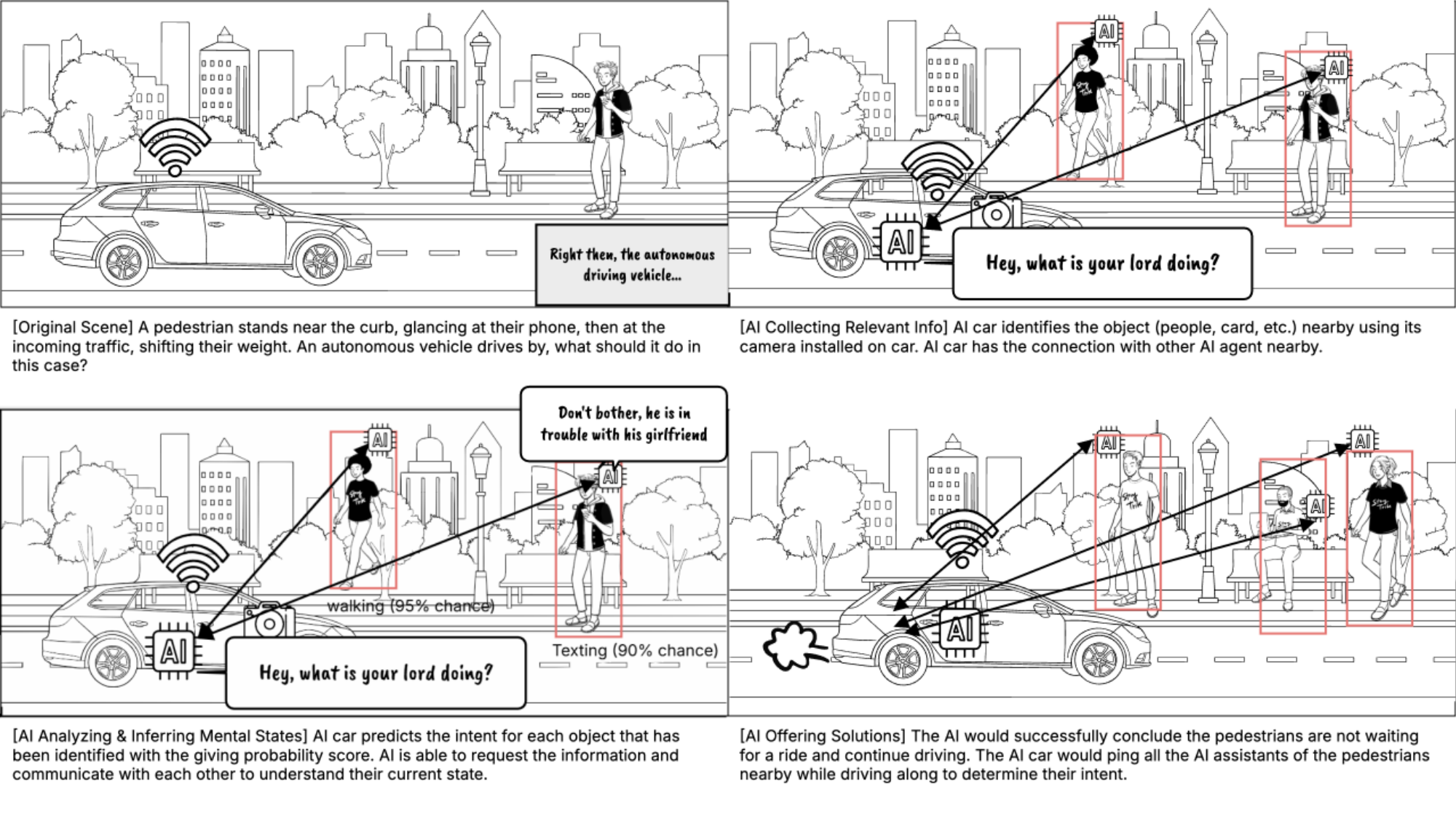}
    \caption{S4 Storyboard created for Autonomous Vehicle scenario. Created with StoryTribe.com.}
    \label{fig:av_storyboard}
\end{figure*}

We found that many practitioners envisioned an IoT-like infrastructure of data collection and communication through existing devices to support the situatedness of ToM-enabled AI, avoiding the need to introduce additional equipments.
% To support the situatedness of ToM-enabled AI, many practitioners envisioned an IoT-like infrastructure of existing devices to collect data and communicate, avoiding the need to introduce additional equipment. 
D5 said, \textit{``I wonder if it's a little more feasible if it (ToM-enabled AI) just connects to a lot of existing devices. So there's no sensor for blood pressure now, but if it just connects to your Apple watch, it automatically gets that information or some other fitness data.''} Both storyboards created by S9 and S13 for the Family AI Assistant scenario revealed practitioners' choices of using small, static devices placed around the house to form a network of devices to detect historical damages, which could be used by the ToM-enabled AI to infer whether the son breaking his mom’s china was intentional or accidental. 

Besides communications between physical devices, many practitioners also proposed the idea of \textit{AI agent communication network} to better infer user's mental states. 
% practitioners who worked on other scenarios also leveraged the idea of ToM-enabled AI communicating with other AI assistants to better infer user’s mental states. 
S1 illustrated in their AI Companion scenario that the ToM-enabled AI robot could ask professional mental health AI agents for resources and advice to deal with the situation. In S4’s storyboard (see Fig.~\ref{fig:av_storyboard}) created for the autonomous vehicle scenario, they envisioned a network of AI agents where the autonomous vehicle could communicate with pedestrians’ personal AI assistants on their smart phones to infer pedestrians’ intention to cross the road. As E4 described during storyboarding: \textit{``We can imagine a whole AI system with different agents connected to each other. For example, a vehicle might detect that another AI agent (of a pedestrian) is active and analyzing direction. Using Bluetooth, they could connect and share information about people’s intentions and actions to create an AI communication network.''}

\subsubsection{Aligning AI Actions with Social Norms and Situational Contexts}
We found that practitioners considered AI acting on mental state inferences as inherently social, hence important to design ToM-enabled AI actions that are aligned with human social norms and contexts. This is especially pronounced in scenarios that consisted of multiple characters such as the AI Trip Planning Assistant and the AI Cooking Robot scenario. For example, S6 (AI Trip Planning Assistant) designed their AI to infer the character’s reluctance to join the trip by drawing on their personal medical information. Their final design showed the AI suggesting alternative plans to better fit the character’s medical needs. When reflecting on their designs, practitioners discussed the complex social norms and consequences that need to be considered in this case:

\begin{quote}
\textit{D6: ``So it's like the transparency isn't there [if the AI doesn't explain why it suggested alternative plans], and it's like how do you make this transparent without giving away that the rationale behind this is because you've inferred a pain state in someone? It's like the AI would have to lie in a way in order to be transparent. It would have to come up with a different reason [for why Lily is reluctant to go on the trip].'' \\
E6: ``So it has to hide this decision making process.'' \\
D6: ``A different, more acceptable decision making process, which is an issue in and of itself. It's an AI trust issue if it's saying that it's making a decision on something that it's actually not.''} 
\end{quote}

Similarly, when designing the AI Cooking Robot scenario, S7 discussed the idea of a ToM-enabled AI that could provide help without disturbing intimate moments:
\begin{quote}
\textit{E7: ``So two different sorts of actions: If Jamie is actually confused and needs help, then the robot should sort of prompt to help. If actually this is an intimate moment between them, then you don't want a robot jumping in the middle.'' \\
D7: ``I wonder if there's some hint to kind of nudge him in the right direction if he actually needs the help.''
E7: ``So maybe a chime, just a noise that it makes to ask, `do you want me to act or not?' That way Jaime can say, `Hey AI, help me.' Instead of just the robot randomly coming in out of nowhere and scaring Sarah.'' }
\end{quote}

Beyond social norms in multi-person settings, we found that practitioners also considered aligning AI actions with situational contexts that shape people's interaction experiences, even in the current dominant one-to-one human-AI interaction paradigm.Practitioners pointed out that many current AI products often treat user requests as isolated one-shot tasks, overlooking users' emotional states, moods, or broader circumstances that could come to play during the interaction. For example, E6 shared how ChatGPT failed at adapting to his emotional needs due to lack of contexts:~\textit{``I use AI in work contexts, helping me code, helping me run through ideas for work, and talk about projects from a student perspective trying to learn things. And ChatGPT… it suffers because it can't see my face, it doesn't do well at assessing my mental state, which has been an emotional thing for me when I'm frustrated because it hasn't helped me get to a specific answer that I need or a goal that I need to get to. So I think ToM could help out with that.''} Similarly, D5 reflected that failure for AI to account for user's subtle cues during interactions could negatively impact user experience:
\begin{quote}
~\textit{``I feel like sometimes people have trouble getting the AI to give them what they want or what they need. Or if they try and interact a certain way with the AI, its tone changes to something they don't like or they just don't like the response in general. It really impacts their experience with it. And so if the AI does have ToM and is able to read into a lot of these, the details and the nitty gritty bits of human behavior and attitude, I think that would really help the AI respond more intelligently and get a little bit closer to what the user is truly trying to get from their experience.''} (D5)
\end{quote}

\subsection{Designing ToM-enabled AI to Support Dynamic Mental States}
We found that practitioners envisioned ToM-enabled AI as capable of adapting to user's fluid and evolving mental states over time, rather than responding only to user's immediate inputs. Practitioners' designs revealed an emphasis for ToM-enabled AI to have ongoing ``awareness'' through continuous monitoring, the need for careful negotiations on the monitoring boundaries with the users, and responsiveness to dynamic mental state changes throughout an interaction.

\subsubsection{Designing for Fluid and Ongoing ``Awareness''}
Our analysis of practitioners' designs surfaced the need for ToM-enabled AI to have fluid and ongoing ``awareness'' of changes in the individuals and the environment over time. This need was demonstrated through participants' design choices in having ToM-enabled AI to collect and analyze relevant longitudinal data of the person as well as continuous monitoring of people's environments to better recognize and predict their real-time mental states. 

Many storyboards generated proposed collecting various historical data to better infer the character’s state of mind in the specific scenarios. Some practitioners brought up the idea of identifying behavior or routine anomalies that were different from the character’s historical pattern. For example, in the AI Companion scenario, S1 and S3 collected historical data such as the character's prior behaviors when entering their home and previous interaction tones with the AI, then compared that against the character's real-time behavior and routine to infer their mental state. Other practitioners fed the ToM-enabled AI with a variety of personal data from outside of the scenario to predict and make sense of the character’s mental states. In the Smart Home Assistant scenario, S10 and S12 designed the character’s smartwatch to automatically sync with the home assistant upon entry, updating it with data such as heart rate, stress levels, sleep, eating patterns, and daily steps throughout the day.

Practitioners extended this idea of ToM-enabled AI’s ongoing awareness to include monitoring environmental changes that could signal the character’s mental states. This was especially true for scenarios akin to smart homes. For the AI Companion scenario (S1) and Smart Home Assistant Scenario (S5, S10), practitioners designed cameras that could continuously capture the character’s home environment and infer negative mental states from the increased messiness of the character’s home compared to before (e.g., wine bottles laying around, uncleaned dishes, clothes on the floor); S10 also designed the AI to be able to detect unusual environmental noises (e.g., keys clattering against the ground, the character’s deep exhale). For the Family AI Assistant scenario, S9’s storyboard embedded valuable home objects with sensors for the AI to continuously monitor object damages and displacement through object movements over time. D9 commented on the importance of incorporating such continuous monitoring and longitudinal data into ToM-enabled AI: \textit{``We initially talked about ToM mostly in terms of what is the person thinking at this very point of time, and that is exactly what we designed for in this [Family AI Assistant] scenario. But I think this scenario was interesting because the larger context of what may have happened before and how a person's reaction is changing over time is also equally important for this complex decision-making.''}

\subsubsection{Negotiating Boundaries of Continuous Monitoring}
While practitioners saw continuous monitoring as the key to enable AI’s ongoing ``awareness,'' they noted that unlike most existing AI products, such level of data collection does not rely on data that users explicitly provided to the AI, which could raise concerns: \textit{``In our applications that we designed [at work], it's mostly around what the user has explicitly provided to the AI, what they offer based on that data as well, as opposed to what we did today in this exercise from the perspective of ToM, the AI assistant picks the data, not like the user offers to it. So this is proactively picking up data.''} (D10) Practitioners voiced their concerns about user's agency in consenting to AI's continuous monitoring. D8 raised concerns about the lack of mechanisms for users to stop the monitoring, but also believed that it could support user’s goals: 
\begin{quote}
\textit{``I feel like if Lily is my friend, I could just ask her or if I notice something, I could go back to our chat history. I may not have access to Lily's health records and fitness goals and everything like that [in this AI Trip Planning Assistant scenario]. It's scary because there doesn't seem to be a mechanism to let the AI know when to stop and what kind of access it could have. But I think if the goal is to make Lily happy about this trip schedule, I think the AI might achieve that goal. It's just like the way it's doing it, I don't know.''} (D8)
\end{quote}
D6 raised concerns about user acceptance of continuous monitoring devices such as the camera that they used in their storyboard:~\textit{``Even when you're with your friends, you're not always looking at 'em, always assessing what they're feeling. Your gaze is not always on them. It's like having something's gaze always on you assessing. You can feel unnatural and weird, and I think people would want to opt out of that eventually because no one wants to feel like they're living life on film, especially when they're at work.''}

Many practitioners believed that careful understanding and negotiations with the users about their privacy boundaries and perceived trade-offs would be necessary to facilitate continuous monitoring: 
\begin{quote}
\textit{``People have suggested many things in our user interviews, like, Hey I want AI to tell me to do this if I'm feeling this, or it should tell me which scissor I should use when I'm pruning this kind of tree, etc. But again, once people understand that this is the kind of data that they would have to provide, having cameras in home or maybe reading this data from their objects in home, it will get messy. So the trade off that some people are willing to make, it also depends on their AI and digital literacy, is important to explore.''} (D9)
\end{quote}
E4 pointed out that it might be necessary to sacrifice our data to gain conveniences from ToM-enabled AI:~\textit{``To be honest, I think we have to sacrifice, but it will be at a certain stage. Right now we're pretty much sacrificing our privacy to the Big Tech. They know what we're searching on, what we're doing, what our closest friend is doing on Facebook, etc.. But we're exchanging [our privacy] with the conveniences. We can look up information really quickly connecting with people who are far away from us...''} As E11 put very aptly on the tension between continuous monitoring and user acceptance:~\textit{``The AI needs help. And to get that, it needs to see the person and understand them, but at the same time, do the people want to be seen?''}

\subsubsection{Adapting to Dynamic Shifts in Mental States} Through our analysis, we found that practitioners proposed various AI solutions that were responsive to user's dynamic mental states in their chosen scenarios, emphasizing the need to design ToM-enabled AI to be adaptive to changes in people's mental states over time.
% Recognizing the need to design ToM-enabled AI that can adapt to changes in people’s mental states over time, practitioners came up with several AI solutions to respond to user’s dynamic mental states in their chosen scenarios. 
For example, S13 (Family AI Assistant) equipped their ToM-enabled AI with an internal cool-down timer that can check back on the mom after a certain period of time. E13 explained this design:~\textit{``Maybe the AI could have an internal timer to check on people. Instead of responding right away, it could wait a few minutes and then ask, `How is everything going?' or offer alternatives. If I’m very mad, I don’t want to think about it at that moment, but I might be more open after I’ve calmed down.''}

In the Autonomous Vehicle scenario, both S11 and S4 incorporated AI confirmations with either the pedestrian or their personal AI assistant to verify the pedestrian’s intention to cross the road before taking action. While designing their AI solution, S11 considered the possibility that a pedestrian might suddenly change their intent and prepared for this by having the vehicle switch to the outer lane to create distance from the pedestrian and slow down in preparation for an immediate stop. E11 highlighted this possibility: \begin{quote}
\textit{``One more thing that came to mind was even if [the pedestrian is] looking around at the shops and trying to find the right shop and all that stuff, and they weren't initially coming up to the curb to cross the street, what happens if they figure out that they did want to go to this shop that's across the street? And humans being humans don't think about it really and just immediately start taking steps towards the street and crossing the street. So you go from not a threat of just looking at the shops around the street to now you're a threat. You're walking out to the street now you become a pedestrian who wants to cross the street.''} (E11)  \end{quote}

Several practitioners highlighted the timing of AI actions as an important design factor, believing that ToM-enabled AI should proactively respond to users’ dynamic mental states.
% Given that ToM-enabled AI can proactively respond to the user's dynamic mental states, several practitioners considered the timing of AI actions an important factor to consider in their design. 
Practitioners felt that AI responses that are too proactive without any prompting could be perceived as strange and uncomfortable. D9 said,~\textit{``An interesting point is, after how long should an AI family assistant in this case even jump in? Probably it's also scary in some ways... Let's say I'm doing a chore and I have it open on my phone and then the AI assistant starts speaking to me or something like that. It would be super weird, right?''} Some practitioners even proposed that ToM-enabled AI should be able to turn itself off based on the user’s mental states. Such AI non-actions were presented as design solutions for the Smart Home Assistant scenario, where both S12 and S5 designed the AI to detect the character’s need for quiet alone time, and acted on such inferred mental state by turning itself off.

\subsection{Designing ToM-enabled AI to Attune for Subjective Individual Mental States}
Our analysis showed that practitioners envisioned ToM-enabled AI as being able to accommodate individual's highly subjective mental states that are largely shaped by subjective experiences, rather than offering generalized AI solutions across users. This design consideration was distilled through practitioners' reflections on the need to design for nuanced and subjective data needs, to account for uncertainty when inferring implicit mental states, and to grapple with the technical challenge of balancing scalable generalization with individual subjectivity.

\subsubsection{Designing for Nuanced and Subjective Data Needs} 
Collecting and analyzing individual data is common in designing personalized AI experiences. However, several practitioners pointed out that ToM-enabled AI is unique given the need to cater to individual mental states that are highly nuanced and subjective to personal traits and experiences. As E10 emphasized, \textit{``I think that a correct behavior from a ToM AI is unique to the user is something that's very distinct about these applications… If the user is an introvert versus an extrovert, the input signals will have very different meanings. So yeah, it seems like it would have to kind of learn the person a little bit.''} D13 also noted that because people differ in their personal characteristics, it would be difficult to infer mental states from behavioral cues alone: 
\begin{quote} 
\textit{``I think everyone has different nuances. We do have standardized ways to tell if someone is angry or not. One thing is the decibels or body temperature. But everyone can have different decibel ranges in terms of when someone is mad. So one person can really be yelling or shouting when they're angry versus another person tends to be more calm when they're angry. I think the way to tell these mental states is very difficult based on certain words or something like that.''} (D13)
\end{quote}

This understanding of mental states being subjective was also illustrated by the extent of highly individualized mental states and personal data that practitioners came up with when brainstorming the mental states and AI techniques. For example, practitioners across all sessions brainstormed extremely personal and subjective data needed to infer mental states, such as the character’s personality types, personal albums, car accident history, ethnic and cultural data, health conditions, all in addition to other historical personal data that we mentioned in the earlier sections such as biometrics data, daily routine, behavioral patterns, etc.. In the AI Trip Planning Assistant scenario, some practitioners brainstormed mental states around personal preferences such as the character's fear of a particular mode of transportation. S2 brainstormed AI techniques and data to infer a highly individualized mental state–-- the character's reluctance to go on the trip due to bad memories tied to the destination. To make such inference, they designed the AI with access to personal data that was associated with the destination, such as personal albums with videos or photos of the destination, private messages with friends on social media about the destination, family and friends’ histories associated with the destination. 

This raised practitioners' major concerns about user privacy. E11 reflected on their choice of sharing biometrics data for the Autonomous Vehicle to infer pedestrian's intent to cross the road:~\textit{``I'll just say right off the bat I would be very, very, very cautious about sharing personal information like that. We're talking about broadcasting personal information like spiked heart rate or what I'm doing on my phone that kinds of stuff that might be useful for the car to determine the person's intentions. But it would also have very serious privacy implications even if it is anonymized. Nothing's ever truly, really anonymized.''}

\subsubsection{Designing with Uncertainty for Implicit Mental States}
% Some practitioners pointed out that ToM-enabled AI are not constrained to explicit user input in specific AI applications or user interfaces as current AI applications are typically designed. D13 said, \textit{“One thing that stood out is ToM-enabled AI will be more reactive than usual [AI applications] where we would proactively seek AI’s help in certain situations, instead of having them actually living in our day-to-day environment. It's not like when I say something, an AI would come in to interrupt saying, ‘Hey, do you need my help?’ It's more like we would go to that application to get help. And also [we build AI applications] that people actually visit the site or the application itself to get help from it.”} 
% D5 and E5’s discussion also echoed this point, suggesting that ToM-enabled AI is similar to agentic AI that go beyond explicit user input:
% \textit{D5: “[ToM-enabled AI] reminds me a little of agentic AI where it’s able to act on behalf of the user, but here it’s kind of taking it one step further where it’s reading more into user intent and then acting on behalf of the user. So it’s almost proactively reading what user intent is and then maybe taking actions based off of that. It depends less on direct user input via text or voice.”}
% E5: "I think you are 100\% right because yield input is something very limited. We have to collect relevant information to support the intent detection, which is really hard." \avanita{Flagging this paragraph for review - content seems misaligned with heading}

Recognizing the subjectivity and implicit nature of individual's mental states, several practitioners voiced their concern that ToM-enabled AI may never be accurate in inferring users’ mental states. E7 noted that even humans cannot directly know others’ mental states, highlighting the fundamental challenge of designing ToM-enabled AI: 
\begin{quote}
~\textit{``In a sense, you can think of somebody's mental state as being the hidden part of the system that's actually emitting different behaviors and observations. [...] So this is called the hard problem in neuroscience where we can assume that other people have mental states, but the only person that we know for sure even has a mental state is ourselves. So we have to infer mental states based on the behavior of a person, like their speech patterns, their tone of voice, their gaze, how they're walking, their facial expressions. I mean there's just so many things in history, you look back through their messages, things that they've done in their past to try to get… you almost have to have a whole history of somebody's life to be able to try to infer what their mental state is.''} (E7)
\end{quote}
Practitioners further elaborated that this challenge was compounded by the fact that there could be multiple valid mental states existing at the same time even within each individual. For instance,in the Family AI Assistant scenario, S13 practitioners highlighted that multiple mental states could co-exist when the character said ``this is why we can't have nice things'' --- she could be both upset about losing the china and also mad at herself.
% D13: ~\textit{ I know we wanted to talk about bringing a sense of warning and urgency to the scene and how she repetitively saying, oh we can't have nice things in a house because of this knowledge or maybe her beliefs. Do you think that's right?}
% E13: ~\textit{Yeah, she's both upset about losing the china and also mad at herself.}

Our analysis showed that most practitioners designed the AI solution to be on the ``safer side'' to account for AI's inaccurate inferences due to the subjective nature of individual mental states. For instance, AI solutions such as offering suggestions based on inferred personal preferences or seeking confirmation instead of acting directly were common across all the storyboards generated.
% To account for inaccurate inferences from the subjective nature of individual mental states, \edit{we found that} most participants designed the AI solution to be on the “safer side” such as offering suggestions based on inferred personal preferences and seeking confirmation instead of acting directly. 
For instance, in the AI Trip Planning Assistant scenario, instead of granting the AI direct control over booking itineraries, S2, S6 and S8 all designed the AI to suggest new travel dates, alternative plans, or alternative itineraries based on the AI’s knowledge of the character’s personal schedule and preferences. Similarly, in the Smart Home Assistant scenario, S5, S10, S12 all designed the AI to take subtle background actions tailored to the character’s personal preferences such as playing their favorite calming music or ordering the character’s favorite food to help the character de-stress. Several practitioners pointed out that these suggestive rather than direct actions from ToM-enabled AI was necessary to avoid putting the AI in an unrecoverable state. 
% D11 explained , 
% \begin{quote}
% ~\textit{ “I feel like the other thing you probably have to do is make decisions that are not catastrophic if you predicted it wrong, where it's like, okay, we know that we only can get to a 99\% confidence threshold, let's make a new decision and then we'll make a new prediction based on that decision. But that first decision needs to not put us in a bad state or needs to not put us in a state that's unrecoverable.”}
% \end{quote}

Practitioners reflected on the challenge of designing ToM-enabled AI that could cater to implicit and subjective mental states. Several practitioners pointed out that designing, or even considering, user's implicit mental states was not something they do at their work practices, in which they focused more on addressing specific user problems. D10 said,~\textit{``In our applications that we design, the input does not account for any [user] mental state. We haven't accounted much for it, honestly. It's mostly around what the user has explicitly provided to the AI.''} E2 further echoed this point and questioned whether incorporating ToM-enabled AI features would help fulfill company priorities in making profits through products: \textit{``You need to sell customers a product. You need to make it really easy for them to envision a future with this product. I feel like it's a bit tricky to make a sales pitch for [ToM-enabled AI] as opposed to something concrete like `here's a iPhone 17 pro, it takes better pictures, battery lasts longer.' ''} Practitioners also expressed their uncertainty about \textit{how} to design AI solutions that could tailor to individual mental states. D13 raised the uncertainty of not knowing how the users were going to perceive of such AI behavior:~\textit{``The difficult part is more so of providing that solution. It can probably detect the same thing from that data, but it's just that how we present it to our customers, they may take it well or they might not take it well.''}

% D12 (01:25:10):
% Because we spend a more time, because typically when I design an AI application, we're focusing more on the problem or the solution. So I feel like either I guess more on solutions and that you make a problem for the solution, that's a typical process. But for this, you're starting with the misalignment and then finding a way to infer about the mental state and then what kind of data we can use for inferring the user's mental state. And then that process was very new to me. The design process is very new to me.

% D10 pointed out that designing for user's implicit mental states is not something they are used to do in their work practice,~\textit{``In our applications that we design, the input does not account for any [user] mental state. We haven't accounted much for it, honestly. It's mostly around what the user has explicitly provided to the AI.''}

\subsubsection{Balancing Generalizability with Individual Subjectivity}
In interviews, many practitioners identified the difficulty of balancing current AI’s development paradigm that focuses on scale and generalizability versus ToM-enabled AI’s requirement of attuning to different individual's subjective mental states. D7 reflected:
\begin{quote}
\textit{``I think just understanding someone's feelings, emotions, wants, desires, is very difficult. I was saying earlier, I'm trying to even think of how the engineers would build a product like this and how they would build them at all. Maybe start with the dataset and kind of add that in? Like in different scenarios you map them out, but it's so complicated. A person can respond in so many ways and have so many infinite amounts of thoughts and feelings, emotions. How do you replicate that?''} (D7)
\end{quote}

Several practitioners also believed that designing ToM-enabled AI would require new approaches that emphasize subjectivity instead of objectivity. E6 explained,~\textit{``I think we've gotten AI pretty good at doing objective things, but involving it in a subjective context is different. It is difficult, and I think we'd have to approach that differently. We'd have to approach it from a less mathematical standpoint and more of a psychological [point of view].'' }

D11 echoed similar points and further reflected on designing for this delicate balance between objectivity and subjectivity via adaptive interfaces: 
\begin{quote}
\textit{``In some cases what we're doing is trying to collect enough data that we build something that is sort of robust to different levels of preexisting knowledge. That is different than building it to focus on the mental states of the users. It’s more like can we make it… almost generically accessible enough that people will not run into major issues, versus how we disambiguate what's going on in their head when they encounter this and then make it fit. Doing that really well probably would require some kind of adaptive interface where the way it presents itself is not traditional software where every person gets the same thing. Maybe we're not actually that far out from that, but it does feel like a step change that's needed if we're going to take ToM really seriously in everyday applications.''} (D11)
\end{quote}

Some practitioners viewed this challenge as the motivation for designing ToM-enabled AI’s capability in learning, adapting, and revising its interpretation of the user over time to better infer and attune to users’ subjective needs. When asked about the opportunities of ToM-enabled AI, E13 emphasized on the importance for ToM-enabled AI to learn about the user over time:~\textit{``Everyone is different. If the AI can just overall knows more about the user over time by integrating the AI more into their daily lives, just collect more data over time, and be able to customize based on different users, I think that would be helpful.''}

% ~\textit{“[Perhaps] incorporate more into daily lives? Being able to learn the pattern of certain people, the target user, because as D13 said earlier, everyone is different. So the difference from having an actual human acting there is that the actual human will either be a friend or a family member. They know that target user will, so they know how they react in different scenarios. So if the AI can just overall knows more about that person over time, just collect more data over time and be able to customize based on different target user, I think that would be helpful.”}

\section{Discussion}
Through co-design sessions with AI practitioners, we surfaced three interrelated design recommendations for ToM-enabled AI in everyday life: it should be \textit{situated} in the social context and norms to interpret how people's mental states are shaped by their immediate surroundings; be responsive to people's \textit{dynamic} and moment-to-moment intentions, emotions, and needs, rather than relying on static or pre-specified assumptions; and be attuned to how individual's \textit{subjective} histories, experiences and nuanced expressions that fundamentally shape how mental states manifest and are communicated. Taken together, these recommendations highlight the need to move beyond inference-based, static mental state representations and one-size-fits-all personalization toward systems that are socially grounded, temporally adaptive, and personally sensitive. 

\begin{figure*}[t]
    \centering
    \includegraphics[width=0.65\linewidth]{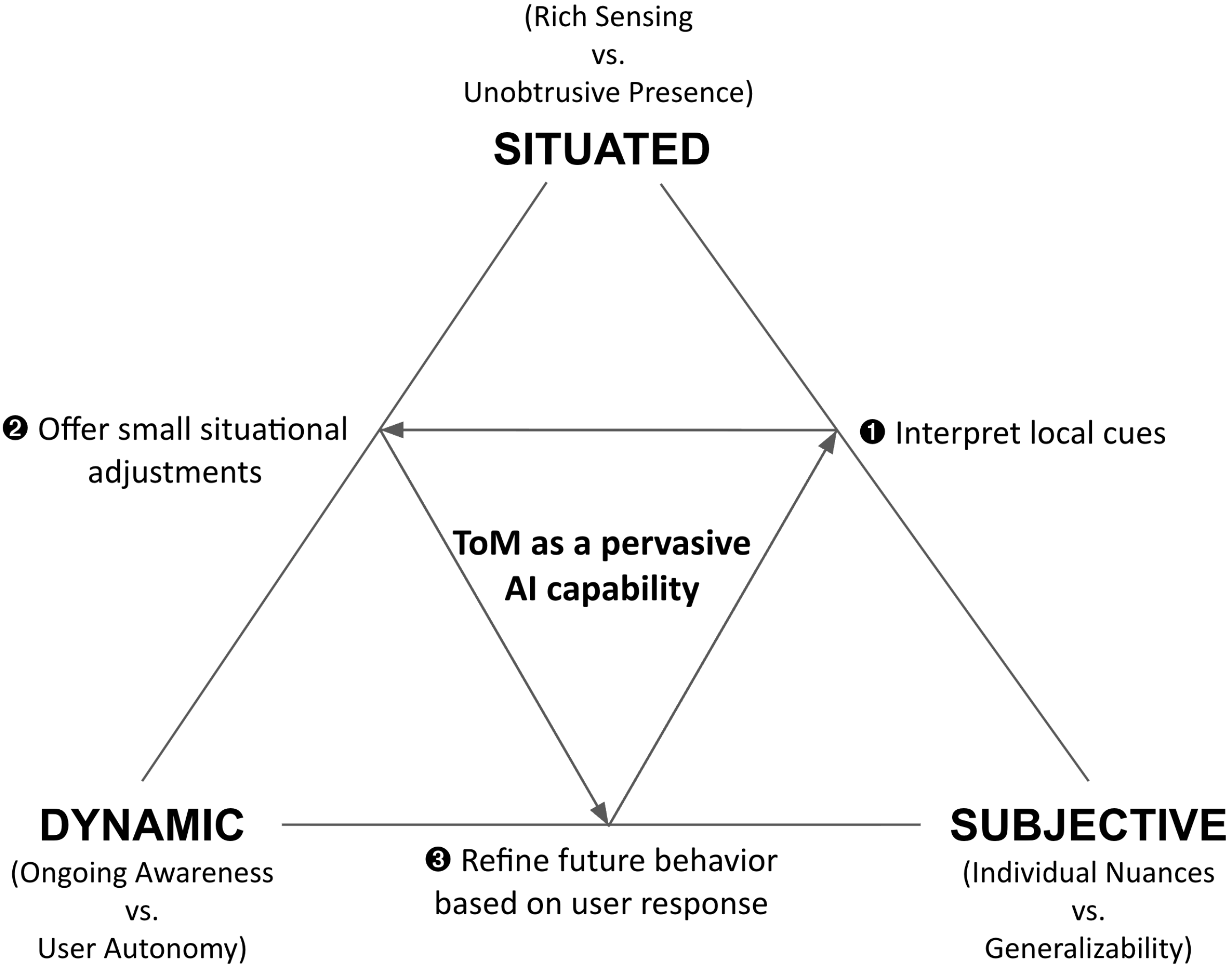}
    \caption{We surfaced three design recommendations for ToM-enabled AI to be situated, dynamic, and subjective, highlighting the need to move beyond inference-based approach to AI ToM in everyday contexts. Each recommendation carries built-in design tensions: ``rich sensing vs. unobtrusive presence'', ``ongoing awareness vs. user autonomy'', and ``individual nuances vs. generalizability.'' We propose the design direction of designing ToM as a pervasive AI capability to enhance intended or existing AI functionalities through \textit{continuous interaction loops} to (1) interpret local cues, (2) offer small situational adjustments, (3) refine future behavior based on user response, and repeat.}
    \label{fig:tom_design_diagram}
\end{figure*}

Compared to previous design research that used similar speculative methods to understand user's perspectives for everyday socially intelligent AI~\cite[e.g.,][]{luria2020social,chang2025unremarkable,reig2021social}, our work offered an opportunity for practitioners to reflect on their designs against the real-world constraints and limitations, surfacing tensions within each design recommendation. These design tensions reveal a broader misalignment between the envisioned future of ToM-enabled AI and the realities of current AI design and development practices, hinting that traditional inference-based approaches to ToM may be insufficient to meet the situated, dynamic, and subjective demands of everyday AI interactions. Below, we build on these insights to reconsider how ToM should be conceptualized in everyday human-AI interactions, examine the tensions and practical constraints that shape its realization, and outline a design direction that treats ToM as a pervasive capability embedded within AI functionalities to support continuous human-AI interaction loops. We summarized these implications in Fig.~\ref{fig:tom_design_diagram}.

\subsection{Beyond Inference: Reframing ToM for Everyday AI}
Our design recommendations for ToM-enabled AI to be situated, dynamic, and subjective highlight that users’ mental states are not static internal variables, but are shaped by context, informed by subjective histories and experiences, and evolve over time. In our study, despite being provided the classical inference-based interpretation of ToM, practitioners consistently imagined far more situated, dynamic, and subjective features for ToM-enabled AI products and services even within short, four-panel storyboards. This demonstrates that socially meaningful human-AI interaction facilitated by ToM-enabled AI requires richer, more contextually grounded forms of understanding than mental state inference alone can provide. 
As such, designing ToM-enabled AI for real-world interactions requires moving beyond the traditional modular and inference-based view of ToM~\cite{baron1985does,gopnik1992child}, which continues to underlie much of today’s AI ToM development and evaluation~\cite{mao2024review} but may not be sufficient to withstand the realities of in-the-wild deployment in everyday user-facing contexts.

This reframing of ToM-enabled AI to be situated, dynamic, and subjective aligns well with the interactionist and enactivist perspectives of social cognition~\cite{de2007participatory,gallagher2008understanding,gallagher2001practice}, which conceptualize social understanding as emerging through ongoing interaction, embodiment, and context rather than as the recognition of static, internal representations. However, compared to the traditional inference-based ToM approach to social cognition, these perspectives could be challenging to operationalize in current AI systems that rely heavily on data, and lack the innate, sophisticated perceptual and interactive capabilities that these theories assume~\cite{de2007participatory,gallagher2008understanding,gallagher2001practice}. Practitioners grappled with this challenge directly in our study: although they envisioned systems sensitive to dynamic contexts, situated subjectivity, and dynamic subjectivity, they also recognized how difficult such visions would be to realize through the data-centric development approach. Challenges such as maintaining extensive context windows, supporting complex data architectures, and addressing privacy and user-acceptance concerns led practitioners, and us, to question whether the dominant data-centric AI development paradigm is sufficient on its own for building situated, dynamic, and subjective ToM-enabled AI. These reflections invite future work to explore participatory sense-making and interactionally grounded approaches to social understanding that can be adapted to the technical and organizational realities of AI development. 

At the same time, we are not arguing for abandoning the inference-based approach to AI ToM altogether: their computational tractability remains valuable and well aligned with prevailing AI paradigms. Perhaps there is an opportunity here to explore hybrid approaches in which inference remains a useful computational tool but is carefully calibrated within the broader interactional processes rather than implemented as a stand-alone module. Additionally, different forms of social understanding may call for different mechanisms--- just as elephants don't play chess~\cite{brooks1990elephants}, ToM-enabled AI may not \textit{always} need such envisioned \textit{full-fledged} situated presence, constant monitoring, and extremely subjective designs to support effective, socially meaningful interactions with humans. Future work should explore how varying degrees of situated, dynamic, and subjective social understanding in AI can also be achieved through ambiguity, intentionally limited inference, or selective visibility (e.g., AI turning itself off).

\subsection{Tensions in Designing ToM-enabled AI on the Ground}
While our findings highlight the importance of designing ToM-enabled AI to be situated, dynamic, and subjective, each of these design recommendations also introduces design tensions (as shown in Fig.~\ref{fig:tom_design_diagram}) that practitioners consistently wrestled with during our study. Designing for \textbf{situated} ToM-enabled everyday AI would require the systems to draw not only on contextual signals facilitated through IoT and ubiquitous sensors~\cite{jaber2024cooking,lim2024exploring,yun2025if,oh2024better}, but also on forms of ``user-awareness'' that capture individual's subtle and subjective cues grounded in those contexts to give mental state inferences valid social meanings. Yet such expanded awareness requires more pervasive and intrusive forms of data collection, placing it in tension with everyday AI's expected non-intrusive presence and alignment with human social norms. The \textbf{dynamic} nature of mental states likewise implies that systems need to monitor change and adapt over time to users' evolving mental states. But the extent of ongoing awareness required for such adaptivity raised significant concerns among practitioners regarding user acceptance, consent, autonomy, and the boundaries of proactive AI behavior—--concerns that are especially fraught in private settings~\cite{luria2020social,lim2024exploring,chalhoub2021did,liao2016can}. This tension consistently appeared in practitioners' designs in installing intrusive data collection devices such as cameras while adding AI mechanisms that can choose to turn itself off. Finally, designing for \textbf{subjectivity} surfaces perhaps the deepest tension: while users and practitioners alike value AI systems that can recognize and adapt to subtle cues and needs that are unique to individual users~\cite{luria2020robotic}, personalization at this level risks collapsing under the modern AI pursuit of generalization and scale~\cite{rafieian2023ai,cunha2018metalearning}.

These tensions further reveal a broader misalignment between the expectations of ToM-enabled AI and the constraints of contemporary AI design and development practices. Practitioners in our study touched upon this misalignment from both the technology-centered and user-centered design perspectives~\cite{yang2020re,yildirim2024discovering}. From a technology-centered design perspective, many questioned the technical feasibility of ToM-enabled AI, pointing to the ethical and legal risks of the extensive data collection required and the limitations of current data architectures and pipelines to process such context-rich, longitudinal, and socially grounded information. Mirroring prior findings on the drawbacks of technology-centered approach~\cite{yang2019sketching}, these concerns would sharply narrow the design space for ToM-enabled AI in real-world AI design and development practice. From a user-centered design perspective, practitioners noted that their design work largely focuses on solving concrete user problems, with limited attention to the mental states users experience while interacting with AI systems. This solution-driven framing can obscure the value of ToM-enabled AI capability, which aims to enhance the quality and nuance of user experience rather than simply optimizing for task completion. At the organizational level, product teams operate under business priorities that privilege scalability, generalizability, and near-term product value~\cite{wang2023designing,yildirim2022experienced}, making it difficult to justify investments in socially intelligent AI capabilities that do not map cleanly onto existing performance metrics or short-term return on investment. Together, these observations suggest that ToM-enabled AI does not fit neatly within current data-driven, solution-driven, and profit-driven development paradigms. This raises a broader question of how such capability could be integrated into everyday AI products and services.

\subsection{Designing ToM-enabled Everyday AI: From Standalone Feature to Pervasive Capability}
These conceptual reframing and tensions motivated us to argue for a different design direction for ToM-enabled AI—--one that treats social understanding not as a stand-alone feature, but as a \textbf{pervasive capability that subtly shapes how AI systems perceive and adapt within the boundaries of their intended or existing functionality}. Embedding ToM within the features that already support core user problems allows product teams to enhance user experience without disrupting the profit-driven or solution-driven priorities of contemporary AI development, while also encourage teams to derive richer social cues from signals systems already encounter rather than hunting for new data or install surveillence mechanism~\cite{cristani2013human}. This design direction also requires structuring the human-AI interaction dynamic as \textit{continuous interaction loops} (as shown in Fig.~\ref{fig:tom_design_diagram}), in which systems interpret local cues, offer small situational adjustments, and refine their behavior over time based on user response. This orientation resonates with emerging interaction frameworks such as Mutual Theory of Mind~\cite{wang2021towards,wang2022mutual}, bi-directional alignment~\cite{shen2024towards}, and socially enactive cognitive systems~\cite{kahl2023intertwining}, all of which view social understanding as iteratively refined through ongoing interaction. For instance, a smartphone assistant noticing abrupt interaction patterns (e.g., repeated shaking or rapid task switching) might infer momentary frustration, offer a small situational adjustment (e.g., prompting ``do you need help''), record and evaluate user's response to such adjustments, and calibrate future behavior based on user feedback. Such contained interaction loops illustrate how situated, dynamic, and subjective forms of ToM can emerge without continuous monitoring or broad data access, aligning with practitioners’ vision of systems that learn and adapt to individual user's social nuances over time.

Treating ToM as a capability also raises methodological challenges for current technical, design, and UX research practices, inviting new approaches that move beyond modular and inference-based approaches to social understanding when designing human-AI interaction. Building systems that interpret and adapt to subtle cues in real time requires methods and techniques that help AI recognize locally meaningful signals and evolve alongside individual users, pushing algorithms designed for population-level generalization towards more personalized trajectories. Yet creating such subjective experience also introduces profound output complexity, especially for more general-purpose AI systems such as smart home assistants. This complicates the use of conventional design methods like sketching, prototyping, or Wizard-of-Oz that assume bounded outputs~\cite{yang2019sketching,yang2020re}, necessitating new design methods and tools that help teams rapidly explore, iterate and evaluate socially nuanced and dynamic AI actions across contexts. 

At the same time, user needs, preferences, and privacy boundaries remain central to making ToM-enabled systems viable. These design requirements are highly subjective and contextualized, hence product teams may need lightweight but situated ways of engaging with users, such as contextual inquiry, diary studies, or experience sampling, to understand how different people negotiate the trade-offs of ToM-enabled AI in everyday settings. This also includes clearer, more granular forms of consent that help users understand what kinds of social cues may be inferred and under what conditions. In this view, ToM-enabled AI becomes a set of bounded, purposeful enhancements to everyday interaction, enriching user experience while remaining feasible, respectful, and aligned with real-world development practices. We encourage future work to further examine this design direction in collaboration with AI practitioners, and surface additional challenges and opportunities to realize ToM as a practical capability in everyday products and services.

\section{Limitations and Future Work}
Our study provides early design recommendations for ToM-enabled AI in everyday contexts, but several limitations shape how our findings and implications should be interpreted. First, our study is grounded in the classic view of ToM that emphasizes on inference-based approach to social understanding, which helped anchor practitioner discussion and designs, but also oriented attention toward certain kinds of social reasoning over others. Future work could explore how alternative perspectives on social understanding, such as embodied, interactional, and participatory accounts, might surface different design considerations. Second, the six human-AI social misalignment scenarios functioned as probes and reflect a narrow slice of everyday life shaped by the U.S.-based everyday AI usage reports and research team experiences. They were useful for eliciting concrete designs but do not capture the full cultural or contextual diversity of everyday human–AI interactions. Expanding scenario settings, sociocultural worlds, and types of ToM-enabled AI behaviors that goes beyond the western, consumer-based, and individualistic scenarios presented in our study would help clarify where our findings and insights do and do not apply. Finally, because our study focused on consumer-facing everyday AI systems, additional work is needed to understand how ToM-enabled AI capability should be designed in other types of everyday AI systems, as well as professional or high-stakes contexts (e.g., ToM-enabled AI in business meetings or healthcare contexts) where expectations, risks, and opportunities may differ.

\section{Conclusion}
In this paper, we conducted 13 co-design sessions with 26 U.S.-based AI practitioners across engineering- and design-oriented roles to envision and reflect on how ToM-enabled AI might manifest in everyday product and service design. Analysis of design artifacts and session transcripts revealed three interrelated design recommendations for designing future ToM-enabled AI products and services: ToM-enabled AI should (1) be \textit{situated} in the social context to interpret how people’s mental states emerge from their environments, (2) be responsive to the \textit{dynamic} and moment-to-moment nature of people's mental states, and (3) be attuned to the \textit{subjective} nature of mental states shaped by individuals’ personal traits and histories. Each recommendation revealed embedded design tensions that point to a broader misalignment between practitioners' envisioned future of ToM-enabled AI and the reality of AI design and development practices. These insights underscored the need to move beyond static, inference-based mental state representations and one-size-fits-all personalization. We argue for designing ToM as a pervasive capability that is woven into AI functionalities to support continuous human-AI interaction loops rather than operating as a discrete, stand-alone module.

% (1) be situated in the social context, (2) be responsive to dynamic human mental states, (3) be attuned for subjective individual differences. These findings led us to reframe ToM-enabled AI from standalone feature to pervasive capability that can greatly enhance human-AI interactions. We discuss the implications of designing for ToM-enabled AI around user acceptance and needs as well as its output complexity.

\begin{acks}
This work was supported by the first author’s Carnegie Bosch Postdoctoral Fellowship. We thank Jessica Lin, Shuhao Ma, Vikram Mohanty, our pilot study participants and those who assisted with study recruitment for their feedback and support. We also thank the anonymous reviewers for their valuable and constructive feedback on earlier drafts.
\end{acks}

% \begin{acks}
% \end{acks}

\bibliographystyle{ACM-Reference-Format}
\bibliography{reference}

\appendix
%TC:ignore

\clearpage
\renewcommand{\thefigure}{\thesection\arabic{figure}}
\renewcommand{\thetable}{\thesection\arabic{table}}
\setcounter{figure}{0}    
\setcounter{table}{0}  

\onecolumn

\section{Reflexive Thematic Analysis} \label{rta}

\footnotesize
\renewcommand{\arraystretch}{1.5}
\begin{longtable}{|p{3.2cm}|p{4.5cm}|p{8.8cm}|}
\caption{\edit{Mapping between design recommendations, themes, and representative example codes from our reflexive thematic analysis of the session transcripts. Only example codes are included here due to space limitations. While the table displays the themes within separate recommendation groups, this structure is simplified for presentation; in our analysis the themes frequently overlapped and contributed to multiple design recommendations.}}\\
\hline
\textbf{Design Recommendation} & \textbf{Themes} & \textbf{Example Codes} \\
\hline
\endfirsthead

\hline
\textbf{Design Recommendation} & \textbf{Themes} & \textbf{Example Codes} \\
\hline
\endhead

\multirow{10}{=}{Designing ToM-enabled AI that is situated in the social context} 
& \multirow{4}{=}{ToM-enabled AI collecting data to infer mental states} 
& ToM-enabled AI data collection equipment can introduce unnaturalness \\
\cline{3-3}
& & ToM-enabled AI can collect data through network of existing devices \\
\cline{3-3}
& & ToM-enabled AI needs to collect data from multiple modalities and sources to infer mental states \\
\cline{3-3}
& & Types of data that can be used to make mental state inferences (e.g., "nonverbal cues and subtle gestures," "need to access massive personal data," "physiological data," "compare and contrast historical data and current state," "user's reactions to the AI can indicate their mental states," "contextual and environmental data.") \\
\cline{2-3}
& \multirow{2}{=}{ToM-enabled AI acting on mental state inferences to align with social norms} 
& ToM AI should be considerate of everyone's mental states in group settings \\
\cline{3-3}
& & ToM AI should be mindful of what information should it disclose to whom \\
\cline{2-3}
& \multirow{2}{=}{Data challenges of designing \& developing ToM-enabled AI} 
& Technical difficulty in designing an architecture to support massive streams of data \\
\cline{3-3}
& & Data are expensive and difficult to obtain, especially during the initial stage of a project \\
\hline

\multirow{14}{=}{Designing ToM-enabled AI to support dynamic mental states} 
& \multirow{3}{=}{ToM-enabled AI acting on mental state inferences to support longitudinal interactions} 
& ToM-enabled AI can correct/revise/update its interpretation of the user based on user feedback over time, rather than fixed assumptions \\
\cline{3-3}
& & AI should anticipate and respond to sudden changes in mental states \\
\cline{3-3}
& & Timing of ToM-enabled AI responses matters \\
\cline{2-3}
& \multirow{2}{=}{Concerns of user losing control over AI's collection and sharing of information} 
& AI sharing information to others \\
\cline{3-3}
& & AI acting on mental state information without user control \\
\cline{2-3}
& \multirow{3}{=}{Concerns of user consent and acceptance} 
& Users might not be comfortable/accepting of the passive and invasive forms of data collection (e.g., camera always collecting and analyzing) \\
\cline{3-3}
& & Challenge and importance of getting user consent on data access \\
\cline{3-3}
& & People might not want their mental states to be known to the AI or other people \\
\cline{2-3}
& \multirow{3}{=}{Tradeoffs between cost and benefit of designing and developing ToM-enabled AI features} 
& How to balance user privacy and the amount of data ToM-enabled AI requires \\
\cline{3-3}
& & Concerns of using ToM-enabled AI might be bigger than the use cases \\
\cline{3-3}
& & Risk of AI misinterpreting users outweighs the convenience of ToM-enabled AI \\
\hline

\multirow{13}{=}{Designing ToM-enabled AI to attune for subjective individual mental states} 
& \multirow{3}{=}{Technical challenge in ensuring the accuracy of ToM inferences} 
& People have very nuanced mental states and behaviors, difficult to generalize \\
\cline{3-3}
& & Physiological data can be up to interpretation when inferring mental states, data can represent multiple valid mental states \\
\cline{3-3}
& & People's behaviors may not always truly represent their mental states. \\
\cline{2-3}
& \multirow{3}{=}{ToM-enabled AI doesn't fit neatly into the current tech design and development paradigm} 
& Practitioners are used to design for explicit user problems and commands instead of implicit and nuanced mental states \\
\cline{3-3}
& & Practitioners are skeptical on whether ToM-enabled AI products/features can drive company profit \\
\cline{3-3}
& & Practitioners work within constraints of data availability \\
\cline{2-3}
& \multirow{3}{=}{ToM-enabled AI making inferences about human mental states} 
& Techniques to engineer ToM-enabled AI seems technically plausible currently \\
\cline{3-3}
& & ToM-enabled AI should infer mental states through probability distribution, which is different than the current classifications and deterministic models \\
\cline{3-3}
& & We need both generalized solution that can be scaled but also personalized solution to enable ToM-enabled AI \\
\cline{2-3}
& \multirow{3}{=}{Opportunities of ToM-enabled AI Applications} 
& To design adaptive AI responses tailored to user's transient mental states, improve UX of AI systems \\
\cline{3-3}
& & ToM-enabled AI can better ensure responses are accurate and adaptive to what the user needs, instead of random guessing \\
\cline{3-3}
& & ToM AI can proactively respond to user's transient mental states like agentic AI \\
\hline

\end{longtable}

\clearpage
\section{Affinity Diagramming} \label{affinity_analysis}
\normalsize
\edit{We used affinity diagramming to cluster the sticky notes participants generated in activity 1 and 2, and the storyboards. The tables below show the example themes and subthemes, and sample data. Note that we also connected sticky notes across the activities for better analytical context--- for example, we connected the AI techniques generated in activity 2 to infer specific mental states in activity 1 during our analysis. We are not able to represent those analytical nuances in these tables due to space constraints.}

\subsection{Activity 1. Brainstorming Mental States}

\footnotesize
\begin{longtable}{|p{3.0cm}|p{4.2cm}|p{7.8cm}|}
% \caption{We used affinity diagramming to cluster the sticky notes about mental states that practitioners brainstormed during the design activity. This table shows the example themes and subthemes, and sample data.}\\
\hline
\textbf{Example Theme} & \textbf{Example Subtheme} & \textbf{Sample Data (sticky notes)} \\
\hline
\endfirsthead

\hline
\textbf{Example Theme} & \textbf{Example Subtheme} & \textbf{Sample Data (sticky notes)} \\
\hline
\endhead

\multirow{5}{3.0cm}{Emotional mental states} 
& \multirow{3}{4.2cm}{Emotional mental states due to situation} 
& [S13] Alice feels upset because she lost her favorite china \\
\cline{3-3}
& & [S7] he's confused about how to cook the dish. \\
\cline{3-3}
& & [S4] Passengers in the car can be frustrated bc they need to get to work \\
\cline{2-3}
& \multirow{2}{4.2cm}{Hidden and implicit emotions} 
& [S6] Lily might have past experience with Miami that she hates it. Trauma associated with Miami. AI suggestion is not vibing with her \\
\cline{3-3}
& & [S8] she is not comfortable speaking up about her worries because jack and amy are really excited. \\
\hline

\multirow{4}{3.0cm}{Intentions} 
& \multirow{2}{4.2cm}{About situation} 
& [S4] Pedestrian is waiting for his car (automatic vehicle), just not that one. \\
\cline{3-3}
& & [S3] Peter is testing the AI assistant \\
\cline{2-3}
& \multirow{2}{4.2cm}{For the future} 
& [S11]A friend of the pedestrian is going to drive by and hand them something--maybe they left the sweater at their friend's house. \\
\cline{3-3}
& & [S13] she wants to warn her son to be careful with nice things at home \\
\hline

\multirow{4}{3.0cm}{Knowledge} 
& \multirow{2}{4.2cm}{Lack of knowledge about self/situation} 
& [S8] she doesnt know anything about the location - she is skeptical about where they are going \\
\cline{3-3}
& & [S7] If its their first date Sarah wont know if Jamie know how to cook or not \\
\cline{2-3}
& \multirow{2}{4.2cm}{Lack of knowledge about AI action/solution} 
& [S6] Jack and Amy don't know much about AI so they just accept AI's solution. \\
\cline{3-3}
& & [S9] husband doesnt know what the future of the gift would look like (high probability of it being broken by the son) \\
\hline

\end{longtable}

\subsection{Activity 2. AI Techniques Brainstorming}

\footnotesize
\begin{longtable}{|p{3cm}|p{4cm}|p{9cm}|}
% \caption{We used affinity diagramming to cluster the sticky notes about AI techniques that practitioners brainstormed to infer specific mental states that they came up with in activity 1. This table shows the example themes and subthemes, and sample data. Note that we also connected sticky notes across the activities for better analytical context--- in our analysis here, we connected the techniques used to infer the specific mental states participants came up with in the previous activities during our analysis. We are not able to represent those analytical nuances in this table due to space constraints.} \\
\hline
\textbf{Example Theme} & \textbf{Example Subtheme} & \textbf{Sample Data (sticky notes)} \\
\hline
\endfirsthead

\hline
\textbf{Example Theme} & \textbf{Example Subtheme} & \textbf{Sample Data (sticky notes)} \\
\hline
\endhead

\multirow{8}{3.0cm}{What data can AI collect}
& \multirow{2}{4.2cm}{Visual cues}
& [S2] AI can monitor lily's facial expression, not maintaining eye contact, eyes dodging \\
\cline{3-3}
& & [S13] There could be a camera that could detect if something similar has happened before \\
\cline{2-3}
& \multirow{2}{4.2cm}{Voice and speech cues}
& [S9] AI can identify Alice's tone and speech, voice data \\
\cline{3-3}
& & [S10] AI can detect unusual loud noises (key clutter, drop noise) compared to normal days \\
\cline{2-3}
& \multirow{2}{4.2cm}{Physiological/biometric information}
& [S7] Heart rate and other sign of nervousness \\
\cline{3-3}
& & [S12] AI could also monitor her temperature to gauge her mental state \\
\cline{2-3}
& \multirow{2}{4.2cm}{Personal data}
& [S7] AI can track whether Jaime made google searches on Sarah to figure out what she likes. \\
\cline{3-3}
& & [S2] AI has access to your personal albums, Ai recognize videos and photo of that place \\
\hline

\multirow{8}{3.0cm}{How can AI make inferences} 
& \multirow{2}{4.2cm}{Multiple sources} 
& [S11]Ride share apps have designated pick-up stops. Check and see if where the pedestrian is standing is near the pick-up spot. Location of the pedestrian (AV can collect data from other apps ) \\
\cline{3-3}
& & [S9] Every object at home is connected in some ways and AI can identify if there is frequent displacement from their existing place - IoT \\
\cline{2-3}
& \multirow{2}{4.2cm}{AI inferring social patterns} 
& [S2] AI could track lily's history when engaging with these events, she's always shy or doesn't like to speak up. \\
\cline{3-3}
& & [S8] AI can know the relationship dynamics between the group \\
\cline{2-3}
& \multirow{2}{4.2cm}{Track behavioral patterns longitudinally} 
& [S3] AI can recognize how frequently Peter talks to the AI assistant \\
\cline{3-3}
& & [S1] Previous recordings would be useful. Interesting and difficult part. Instead of judging current state, the AI should infer and gain previous learnings about the person's personality. \\
\cline{2-3}
& \multirow{2}{4.2cm}{AI self monitoring} 
& [S5]AI just turns itself off for the night. Lina starts walking up to the AI in an angry manner... \\
\cline{3-3}
& & [S4] AI should provide rationales of its abnormal AI behavior. \\
\hline

\end{longtable}
\vspace{10mm}

\subsection{Activity 3. Storyboarding}

\footnotesize
\begin{longtable}{|p{3.0cm}|p{4.2cm}|p{7.8cm}|}
\hline
\textbf{Themes} & \textbf{Example Subthemes} & \textbf{Sample Data (Researcher Notes/Observation)} \\
\hline
\endfirsthead

\hline
\textbf{Themes} & \textbf{Example Subthemes} & \textbf{Sample Data (Researcher Notes/Observation)} \\
\hline
\endhead

\multirow{3}{3.0cm}{AI forms} 
& \multirow{2}{4.2cm}{Home decor} 
& [S5] Photo frame on the wall \\
\cline{3-3}
& & [S7, S10, S12] Alexa / Nest /tablet \\
\cline{2-3}
& Robotic AI assistant 
& [S13, S1, S4] \\
\hline

\multirow{9}{3.0cm}{AI data collection} 
& \multirow{2}{4.2cm}{Continuous data collection} 
& [S13] collect historical data \\
\cline{3-3}
& & [S2] collect digital footprint - browsing history content \\
\cline{2-3}
& \multirow{3}{4.2cm}{Multimodal data collection} 
& [S8] calendar + personal communications + personality + physiological data \\
\cline{3-3}
& & [S4] facial expression + posture + gaze direction -> from others phones / services \\
\cline{3-3}
& & [S5] Camera + sensors - images/ visual information, detect things on the flooe \\
\cline{2-3}
& \multirow{2}{4.2cm}{Connected devices} 
& [S4]data type - from other AI agents \\
\cline{3-3}
& & [S9]data type - Sensors for interconnected devices \\
\hline

\multirow{5}{3.0cm}{AI inferring mental states} 
& \multirow{2}{4.2cm}{Comparing historical data against current} 
& [S13] comparing historical behaviour with current behaviour to get mental states \\
\cline{3-3}
& & [S1]check delta as compared to previous behaviours \\
\cline{2-3}
& \multirow{2}{4.2cm}{Most probable cause of mental state} 
& [S8] highest most likely issue - probability - output as JSON file \\
\cline{3-3}
& & [S4]most probable action that AV thinks is applicable \\
\hline

\multirow{11}{3.0cm}{AI solutions} 
& \multirow{3}{4.2cm}{Confirming before acting} 
& [S3] confirm inclination and action \\
\cline{3-3}
& & [S11]confirming - ask pedestrian for clarification \\
\cline{3-3}
& & [S1] ask what it can do to cheer up \\
\cline{2-3}
& \multirow{3}{4.2cm}{Offering suggestions} 
& [S13] Suggestive - giving son advice and emotional supprt \\
\cline{3-3}
& & [S2] suggestive - proposes new date after looking at calendars of everyone in group \\
\cline{3-3}
& & [S10] suggestive - offers lina suggestions for her favourite food and tv show \\
\cline{2-3}
& \multirow{3}{4.2cm}{Subtle background actions} 
& [S5] corrects action - stops coffee machine and energetic music \\
\cline{3-3}
& & [S12]AI stops interactions with lina \\
\cline{3-3}
& & [S5] goes to sleep to avoid intrusion and offer space \\
\cline{2-3}
& \multirow{2}{4.2cm}{Invasive actions} 
& [S1] play fav music, tell a joke, dance \\
\cline{3-3}
& & [S11] change lanes \\
\hline

\end{longtable}

% \appendix




\end{document}
\endinput
%%
%% End of file `sample-sigconf-authordraft.tex'.